# Incorporating the effect of white matter microstructure in the estimation of magnetic susceptibility in ex vivo mouse brain

Anders Dyhr Sandgaard[1], Valerij G. Kiselev[2], Rafael Neto Henriques[3], Noam Shemesh[3], Sune Nørhøj Jespersen[1,4]

[1]Center for Functionally Integrative Neuroscience, Department of Clinical Medicine, Aarhus University, Denmark

[2]Division of Medical Physics, Department of Radiology, University Medical Center Freiburg, Freiburg, Germany

[3]Champalimaud Research, Champalimaud Centre for the Unknown, Lisbon, Portugal,

[4]Department of Physics and Astronomy, Aarhus University, Denmark

**Word count:** 4491 (main body)

**Corresponding Author**: Sune Nørhøj Jespersen.

**Mail**: sune@cfin.au.dk

# Abstract

**Purpose:** To extend Quantitative Susceptibility Mapping (QSM) to account for microstructure of white matter (WM) and demonstrate its effect on ex vivo mouse brain at 16.4T.

**Methods:** Previous studies have shown that the MRI measured Larmor frequency also depends on local magnetic microstructure at the mesoscopic scale. Here, we include effects from WM microstructure using our previous results for the mesoscopic Larmor frequency $\overline{\Omega}^{\text{Meso}}$ of cylinders with arbitrary orientations. We scrutinize the validity of our model and QSM in a digital brain phantom including $\overline{\Omega}^{\text{Meso}}$ from a WM susceptibility tensor and biologically stored iron with scalar susceptibility. We also apply susceptibility tensor imaging (STI) to the phantom and investigate how the fitted tensors are biased from $\overline{\Omega}^{\text{Meso}}$. Last, we demonstrate how to combine multi-gradient echo (MGE) and diffusion MRI (dMRI) images of ex vivo mouse brains acquired at 16.4T to estimate an apparent scalar susceptibility without sample rotations.

**Results:** Our new model improves susceptibility estimation compared to QSM for the brain phantom. Applying STI to the phantom with $\overline{\Omega}^{\text{Meso}}$ from WM axons with scalar susceptibility produces a highly anisotropic susceptibility tensor that mimics results from previous STI studies. For the ex vivo mouse brain we find the $\overline{\Omega}^{\text{Meso}}$ due to WM microstructure to be substantial, changing susceptibility in WM up to 25% root-mean-squared-difference.

**Conclusion:** $\overline{\Omega}^{\text{Meso}}$ impacts susceptibility estimates and biases STI fitting substantially. Hence, it should not be neglected when imaging structurally anisotropic tissue such as brain WM.

# Introduction

Quantitative Susceptibility Mapping[1–4] (QSM) is a commonly utilized MRI methodology for mapping tissue susceptibility[5]. Its application in disease is highly promising for imaging changes in tissue iron, calcium and myelin[6–9]. Voxel-specific tissue magnetic susceptibility can be estimated from the gradient-recalled echo (GE) signal phase by inverting the measured magnetic field offset as a simple Fourier space product of the main-field induced magnetization with the Lorentz-corrected dipole kernel[10]. This relation holds however in general only for isotropic media with scalar susceptibility.

One of the shortcomings of the current QSM framework is the neglect of mesoscopic field effects associated with microstructure and anisotropic susceptibility. This assumption is especially challenged in white matter (WM) tissue, where field perturbations from WM axons have been observed to depend on the orientation to the external field[11–16].

A measurable orientational dependence of the magnetic field – here termed *magnetic anisotropy* – may originate from different underlying length scales. On the macroscopic scale, the overall sample shape gives rise to an orientation dependent field - including the effect of multiple tissue regions with different magnetic properties such as WM and gray matter. Such types of magnetic anisotropy are already considered in QSM or Susceptibility Tensor Imaging[17] (STI), which extends QSM to a tensor valued susceptibility. A measured magnetic anisotropy can also stem from microscopic field effects far below the sampling resolution (sub-voxel). This naturally occurs due to microscopic susceptibility anisotropy, such as the alkyl chains in the myelin sheaths[18–20]. However, anisotropy also arises in systems with only a scalar susceptibility arranged in a microscopically anisotropic structure. We refer to these two distinct origins of magnetic anisotropy as *microscopic susceptibility anisotropy* and *microscopic structural anisotropy*, respectively, to separate from macroscopic effects. Note that macroscopic strategies, such as STI, are affected by both micro- and macroscopic magnetic anisotropy but cannot distinguish between the two, as mesoscopic field effects are unaccounted for in the STI framework. Wharton and Bowtell[21] measured the frequency shift outside a fresh porcine optic nerve, and estimated the contribution from the sample, assumed to have both isotropic and anisotropic susceptibility components, with high precision. They found that the susceptibility anisotropy contributed around 5 times less to the measured frequency shift than the isotropic susceptibility component. This suggests that a minimal extension to QSM that captures magnetic anisotropy should incorporate mesoscopic field effects arising from structural anisotropy but could neglect susceptibility anisotropy to a first approximation. This would also account for the effects of WM orientation dispersion, which can greatly affect mesoscopic frequencies[22] and constitute a substantial part of the total Larmor frequency[21] shift.

Recently, we outlined a framework describing the MRI measured Larmor frequency shift $\overline{\Omega}_{MRI}(\mathbf{R})$[22]. We investigated microstructural effects for a population of long solid cylinders with scalar susceptibility and arbitrary orientation dispersion and found that the mesoscopic contribution depends on $l = 2$ Laplace expansion coefficients, $p_{2m}$, of the fiber orientation distribution function (fODF). These findings bridge the gap between fully parallel and uniformly dispersed cylinders previously used to describe microstructural field effects from cylinders[11–14], without the need to assume a low volume fraction[11].

Here we use this framework to address one of the shortcomings of QSM, namely, the unaccounted for mesoscopic frequency shifts, to present a minimal biophysical model of the MRI measurable Larmor frequency offset. We combine Larmor frequency measurements with fODF information ($p_{2m}$) obtained by Fiber Ball Imaging[23] (FBI) diffusion MRI (dMRI). This enables estimation of the voxel-averaged (bulk) scalar magnetic susceptibility of our model that includes orientation dependent mesoscopic frequency shifts in WM but without the need for imaging at multiple sample orientations. This is different to previous studies[24,25] using information from Diffusion Tensor Imaging[26] (DTI) to estimate the orientation of the STI susceptibility tensors, which neglected any form of structural anisotropy and mesoscopic frequency shifts. To our knowledge, only one previous study[16] has included a mesoscopic frequency shift from axially symmetric WM axons with scalar susceptibility to QSM, where the fitted susceptibility from the standard QSM-part reflected the total bulk scalar susceptibility from WM myelin, iron etc. Estimating both parameters required sample rotations and the orientation dependence was approximated by the primary eigenvector of the DTI diffusion tensor. Corrections due to the local frequency shift from chemical exchange has also been considered previously[27] in QSM.

Here we use the estimated fODF to determine a WM specific local mesoscopic contribution to the MRI Larmor frequency, representing a novel contrast based on combined information from susceptibility and fODF. We argue that this model captures the predominant effects contributing to the measured Larmor frequency shift, equivalent to making the following three approximations (P1-P3), which we justify in the theory section:

- **P1)** Magnetic anisotropy of myelin is mainly caused by microscopic structural anisotropy with the magnetic susceptibility approximated as a scalar.
- **P2)** The variance in the voxel-wise bulk susceptibilities of iron in highly structurally anisotropic WM is less than the variance in bulk susceptibility of myelin.
- **P3)** Additional exchange-related frequency effects in myelin water are subdominant to the total measured Larmor frequency related to susceptibility.

For this, we extend our model[22] for solid cylinders to multi-layered cylinders to describe the mesoscopic frequency shifts from the WM microstructure with approximately scalar susceptibility.

Here, we investigate the parameter accuracy of QSM compared to our new framework by constructing a digital susceptibility brain phantom from dMRI images that includes both isotopic and anisotropic susceptibility of WM, and an iron-related scalar susceptibility in both WM and gray matter (GM). We find that our model improves fitting over QSM as long as the variance of myelin bulk susceptibility is greater than that of bulk WM iron susceptibility. This is also true when the absolute mean of bulk WM iron is lower than the myelin bulk susceptibility. We further simulate the frequency shift acquired at multiple sample directions, where frequency shifts from WM susceptibility anisotropy are turned on or off. By applying STI[17,28,29], we investigate the fitted tensor susceptibility originating exclusively from unaccounted mesoscopic frequency shifts from the WM microstructure with only scalar susceptibility. This reveals a major bias in the apparent susceptibility tensor from microscopic structural anisotropy, which turns out to be much greater than the effect from actual susceptibility anisotropy (microscopic and macroscopic). Last, we apply our model framework for the frequency shift $\overline{\Omega}_{\text{MRI}}(\mathbf{R})$ to experimental MGE and dMRI data obtained in ex vivo mouse brain. We estimate the voxel averaged Larmor frequency, and show that mesoscopic frequency shifts can be of the same order of magnitude to the measured frequency shift, and change susceptibility estimation in highly structural anisotropic WM.

## Theory

We start by outlining the considered system, along with a brief summary of the framework for the MRI measured position-dependent Larmor frequency[22] $\overline{\Omega}_{\text{MRI}}(\mathbf{R})$ based on the principle of coarse graining and by using a mesoscopic Lorentz sphere construction[30–33]. Then we extend our solution for the Larmor frequency from infinite solid cylinders with arbitrary orientation dispersion to include multilayer cylinders as shown in Figure 2.

*System of consideration*

We describe the macroscopic sample of volume V as a porous medium consisting of impermeable microscopic magnetic inclusions, e.g., myelin lipid bilayers. The spatial organization of the inclusions is represented by the microscopic indicator function $v(\mathbf{r})$, which is 1 inside inclusions and 0 otherwise. This defines the *microstructure* (depicted as cylinders in Figure 1D). We assume inclusions are dia- or paramagnetic, and uniformly magnetized along the applied field $\mathbf{B}_0 = B_0\widehat{\mathbf{B}}$, where $\widehat{\mathbf{B}}$ is a unit vector (as are all hatted vectors in what follows). The magnetization is described by a microscopic magnetic susceptibility $\chi(r) \propto v(r)$ being on the order of ppm and given relative to the susceptibility of water (see supplementary material S1 for a detailed description of referencing).

**Modeling a population of multilayered cylinders**

The MRI measured Larmor frequency shift $\overline{\Omega}_{\text{MRI}}(t; \mathbf{R})$ of the gradient-echo signal $S(t; \mathbf{R})$ is perturbed by local magnetic field variations induced by the tissue. Here $\mathbf{R}$ denotes the center of the voxel and $t$ the echo time. As shown in previous studies[31–35], $\overline{\Omega}_{\text{MRI}}(\mathbf{R})$ can be decomposed into two contributions depending on the distance to $\mathbf{R}$ and the echo time $t$ (in the absence of background sources)

$$\overline{\Omega}_{\text{MRI}}(t; \mathbf{R}) = \overline{\Omega}^{\text{Meso}}(t; \mathbf{R}) + \overline{\Omega}^{\text{Macro}}(\mathbf{R}) + \overline{\Omega}_{\text{Ref}}(\mathbf{R}). \tag{1}$$

$\overline{\Omega}^{\text{Macro}}(\mathbf{R})$ captures the time independent frequency induced by distant sources on the macroscale (Figure 1C) and depends on the sample shape,

$$\overline{\Omega}^{\text{Macro}}(\mathbf{R}) = \gamma B_0 \widehat{\mathbf{B}}^T \sum_{\mathbf{R}'} \overline{\mathbf{Y}}(\mathbf{R} - \mathbf{R}') \overline{\chi}(\mathbf{R}') \widehat{\mathbf{B}}, \tag{2}$$

where $\overline{\mathbf{Y}}(\mathbf{R} - \mathbf{R}')$ is the voxel-averaged dipole kernel centered at every sampling position $\mathbf{R}'$. $\overline{\Omega}_{\text{Ref}}(\mathbf{R})$ defines the frequency offset[22] at $\mathbf{R}$ from the chosen reference susceptibility and is removed upon background field removal[36]. Supplementary material S1 explains this referencing in more detail, including a simulation demonstrating the removal of $\overline{\Omega}_{\text{Ref}}(\mathbf{R})$. $\overline{\Omega}^{\text{Meso}}(t; \mathbf{R})$ is a time dependent frequency offset induced by explicit magnetic microstructure in the mesoscopic vicinity of $\mathbf{R}$ (Figure 1D)[37]. When $S(t; \mathbf{R})$ is measured in either the static dephasing regime or diffusion narrowing regime[38] assuming non-exchanging compartments in the diffusion narrowing regime, $\overline{\Omega}^{\text{Meso}}(t; \mathbf{R}) = \overline{\Omega}^{\text{Meso}}(\mathbf{R}) + \mathcal{O}(t)$ is a power law series in time, where the time independent term $\overline{\Omega}^{\text{Meso}}(\mathbf{R})t$ approximates the first signal cumulant for weak dephasing. The first cumulant is convenient as it describes the mean frequency sampled by the point-spread-function. Assuming that the magnetic microstructure varies slowly compared to the imaging resolution, with a locally uniform scalar magnetic susceptibility (as shown in Figure 1A),

$$\overline{\Omega}^{\text{Meso}}(\mathbf{R}) \approx \gamma B_0 \hat{\mathbf{B}}^T \mathbf{L}(\mathbf{R}) \hat{\mathbf{B}}, \quad \text{(Slowly varying microstructure)}. \tag{3}$$

Here $\mathbf{L}(\mathbf{R})$ is the mesoscopic Lorentzian tensor[14,22,35]. For uniform susceptibility $\mathbf{L}(\mathbf{R}) = -\chi(\mathbf{R})\mathbf{N}(\mathbf{R})$ where $\mathbf{N}(\mathbf{R})$ is a mesoscopic demagnetization tensor[22] depending only on structural correlations near $\mathbf{R}$, and $\chi(\mathbf{R})$ is the local magnetic susceptibility of cylinders. We previously derived $\mathbf{N}(\mathbf{R})$ for a population of solid long cylinders exhibiting arbitrary orientation dispersion[22]. In WM fibers, water resides not only outside cylinders, but also in the intra-axonal space and myelin bilayers. In supplementary material S2 we extend our cylinder model to include multilayer cylinders (as shown in Figure 2) and show that $\mathbf{N}(\mathbf{R})$ is in fact identical to the result for solid cylinders. This means that the mean Larmor frequency in any water compartment is indistinguishable from that in any other for this magnetic microstructure. The model-specific MRI Larmor frequency $\overline{\Omega}_{\text{MRI}}(\mathbf{R})$, Equation (1), finally becomes

$$\overline{\Omega}_{\text{MRI}}(\mathbf{R}) = \gamma B_0 \left( -\overline{\chi}(\mathbf{R}) \frac{1}{3} \sum_{m=-2}^{2} p_{2m}(\mathbf{R}) Y_{2m}(\hat{\mathbf{B}}) M^{\text{WM}}(\mathbf{R}) + \hat{\mathbf{B}}^T \sum_{\mathbf{R}'} \overline{\mathbf{Y}}(\mathbf{R} - \mathbf{R}') \overline{\chi}(\mathbf{R}') \hat{\mathbf{B}} \right) + \overline{\Omega}_{\text{Ref}}(\mathbf{R}). \tag{4}$$

The first term in Equation (4) is $\overline{\Omega}^{\text{Meso}}(\mathbf{R})$. Here $\overline{\chi}(\mathbf{R})$ defines the mesoscopically averaged (bulk) magnetic susceptibility, $M^{\text{WM}}(\mathbf{R})$ is a binary mask of WM (not to be mistaken for the magnetization). The orientation dependence is captured by the $l = 2$ Laplace expansion coefficients $p_{2m}(\mathbf{R})$ of the fiber orientation distribution (fODF) measurable with dMRI[23,39,40], and $Y_{2m}$ is the $l = 2$ spherical harmonics. Equation (4) differs from the conventional QSM by the presence of a mesoscopic contribution from local magnetic microstructure[14,22,35], and by using a voxel averaged dipole field $\overline{\mathbf{Y}}$ as opposed to the elementary field[4] $\mathbf{Y}$. We have previously shown with simulations[22] that both can have a substantial effect on estimating Larmor frequencies.

**Frequency contributions from WM susceptibility anisotropy**

The microscopic susceptibility tensor $\boldsymbol{\chi}$ for a single lipid pointing along $\hat{\mathbf{u}}$ constituting the myelin sheet of a multilayer cylinder with axial direction $\hat{\mathbf{n}}$ is

$$\boldsymbol{\chi} = (\chi - \tfrac{1}{3}\Delta\chi)\mathbf{I} + \Delta\chi \hat{\mathbf{u}}\hat{\mathbf{u}}^T. \tag{5}$$

$\Delta\chi$ defines the susceptibility anisotropy along $\hat{\mathbf{u}}$ and $\chi = \text{Tr}[\boldsymbol{\chi}]/3$ is a third of the trace.

Averaging over lipids and cylinders (denoted by $\langle \cdot \rangle$), the bulk magnetic susceptibility $\overline{\boldsymbol{\chi}}$ of many multilayer cylinders with arbitrary orientations is

$$\overline{\boldsymbol{\chi}} = \zeta\langle\boldsymbol{\chi}\rangle = \zeta\left(\chi\mathbf{I} - \frac{\Delta\chi}{2}\left(\mathbf{T} - \frac{1}{3}\mathbf{I}\right)\right) = \left(\overline{\chi}\mathbf{I} - \frac{\Delta\overline{\chi}}{3}\sum_{m=-2}^{2}p_{2m}\boldsymbol{\mathcal{Y}}_{2m}\right). \tag{6}$$

$\zeta$ is the volume fraction of the cylinders. Here we utilized the axial symmetry of the lipids for each multilayer cylinder and $\langle\hat{\mathbf{u}}\hat{\mathbf{u}}^T\rangle = \frac{1}{2}(\mathbf{I} - \mathbf{T})$, where $\mathbf{T} = \langle\hat{\mathbf{n}}\hat{\mathbf{n}}^T\rangle$ is the scatter matrix[41]. Using the relation $\hat{\mathbf{n}}\hat{\mathbf{n}}^T = 1/3\mathbf{I} + 8\pi/15\sum_{m=-2}^{2}\boldsymbol{\mathcal{Y}}_{2m}Y_2^m(\hat{\mathbf{n}})$, where $\boldsymbol{\mathcal{Y}}_{2m}$ are the symmetric trace-free tensors (STF) corresponding to an irreducible rank-2 representation of SO(3)[42], and representing $\langle \cdot \rangle$ as an integral with the fODF[22], the scatter matrix $\mathbf{T}$ could be rewritten in terms of $p_{2m}$, the Laplace expansion coefficients of the fODF $\mathbf{T} = 1/3\mathbf{I} + 8\pi/15\sum_{m=-2}^{2}\boldsymbol{\mathcal{Y}}_{2m}p_{2m}$, leading to the last equality in Equation (6).

The macroscopic contribution $\overline{\Omega}_{\Delta\chi}^{\text{Macro}}(\mathbf{R})$, Equation (2), due to non-zero $\Delta\chi$ is thus

$$\overline{\Omega}_{\Delta\chi}^{\text{Macro}}(\mathbf{R}) = -\frac{1}{3}\hat{\mathbf{B}}^T\sum_{\mathbf{R}'}\overline{\mathbf{Y}}(\mathbf{R} - \mathbf{R}')\Delta\overline{\chi}(\mathbf{R}')M^{\text{WM}}(\mathbf{R}')\sum_{m=-2}^{2}p_{2m}(\mathbf{R}')\boldsymbol{\mathcal{Y}}_{2m}\hat{\mathbf{B}}. \tag{7}$$

Equation (7) gives an explicit description of the dependence of the macroscopic frequency shift on fiber orientation through $p_{2m}$ and susceptibility anisotropy through $\Delta\overline{\chi}$. The mesoscopic contribution $\overline{\Omega}_{\chi}^{\text{Meso}}(\mathbf{R})$ from $\chi$ is found by extending our previous model[22] to multilayer cylinders, cf. Equation (4). However, no analytical results for the mesoscopic contribution $\overline{\Omega}_{\Delta\chi}^{\text{Meso}}(\mathbf{R})$ from orientationally dispersed multilayer cylinders with susceptibility anisotropy $\Delta\chi$ exist. However, as described in previous work[22], it is given by a Lorentzian tensor $\mathbf{L}_{\Delta\chi}$ which depends on a cross correlation tensor $\boldsymbol{\Gamma}^{v\Delta\chi}$ between the reporting NMR-visible fluid and the anisotropic susceptibility.

**Minimal model framework for susceptibility estimation**

It is well known that WM myelin includes susceptibility anisotropy due to lipid chains[18–20], but also contributions from other sources such as iron[43,44]. In addition, a high frequency shift in myelin water is usually ascribed to exchange[13,14,45–47]. Estimating all parameters is a daunting task, especially when mesoscopic frequency shifts must be accounted for, and would generally require active sample rotations, which might not be clinically feasible.

In the pursuit of rotation-free susceptibility estimation, we propose Equation (4) as a minimal biophysical model framework to account only for major susceptibility sources in each voxel. This model includes the mesoscopic frequency shifts from the WM microstructure albeit with scalar susceptibility, and thus neglects susceptibility anisotropy (**P1**) - just like QSM. Neglecting WM susceptibility anisotropy as a first approximation can be justified by a previous study[21] estimating the magnitude ratio between the isotropic and anisotropic parts of WM susceptibility to be around 5:1 with high precision. WM iron, in the region of $M^{WM}$ where we explicitly model susceptibility sources as myelin, is assumed to be uniformly distributed (**P2**). This is justified when the mean magnitude in bulk susceptibility of WM iron is lower than the bulk susceptibility of WM myelin, or when the variance in bulk susceptibility of WM iron is subdominant compared to the variance in bulk susceptibility of WM myelin susceptibility[43] (see simulation, cf. Figure 5). As shown in section 1 in the supplementary material (S1), we can then neglect WM iron susceptibility in $\overline{\Omega}_{MRI}(\mathbf{R})$, as it re-appears in $\overline{\Omega}_{Ref}(\mathbf{R})$ and as a shift in susceptibility in GM and CSF. Then, after estimating the susceptibility and referencing it to the found CSF susceptibility, WM susceptibility represents a sum over iron and myelin bulk susceptibility referenced to CSF. The contribution from myelin water (**P3**) can be disregarded by exploiting its very fast relaxation rate[48], i.e., by estimating the Larmor frequency only at echo times much greater than its relaxation time.

Next, we investigate these assumptions and the parameter accuracy of our framework compared to QSM.

## Methods

**Ex vivo brain imaging**

All animal experiments were preapproved by the competent institutional and national authorities and carried out according to European Directive 2010/63.

*Animal preparation*

Animal experiments were performed on a perfusion-fixed C57Bl6 mouse brain. Briefly, a mouse was euthanized prior to the experiment with pentobarbital, transcardially perfused with phosphate-buffered saline (PBS) followed by a 4% paraformaldehyde (PFA) solution. The brain was then extracted and stored in 4% PFA for about a week in a fridge at 4 degrees Celsius, and 37 degrees one day prior to imaging so the brain could reach thermal equilibrium with the scanner room. Before imaging, the brain was washed with PBS to minimize relaxation-effects induced by the fixative[49]. The brain was subsequently placed axially in a 10 mm NMR tube and filled with Fluorinert (Sigma Aldrich, Lisbon, Portugal).

*MRI experiments*

Experiments were performed on a 16.4 T Bruker Ascend Aeon (Bruker, Karlsruhe, Germany) interfaced with an Avance IIIHD console and a 10 mm Micro5 probe equipped with gradients capable of delivering up to 3 T/m in all directions. Remmi sequences (Remmi) were used to acquire 3D gradient-recalled multi-echo images (MGE) and 3D dMRI images. For all acquisitions, repetition time was kept at 20 ms, flip angle at 20 degrees and bandwidth of 150 kHz. The Field-of-View for these 3D acquisitions was 10.2×17.0 ×10.2 mm$^3$, matrix size 102×170×102 which resulted in an isotropic resolution of (100 μm)$^3$. For MGE, the echo times were 1.75, 3.5,…, 17.50 ms, while dMRI was acquired at 11, 12.55,…, 19.75 ms. Two experiments with four averages were acquired for the MGE leading to an SNR in WM up to 40 and 45 in GM. dMRI was acquired with b-values ranging from 1 to 3 ms/μm$^2$, with 30 directions (**exp1**). In another experiment with identical acquisition parameters, the diffusion parameters were set to b=5 ms/μm$^2$ and 10 ms/μm$^2$ along 75 directions (**exp2**). 1 average was performed for dMRI experiments leading to an SNR in WM up to 15 and 5 in GM for b=5 ms/μm$^2$, and 10 in WM and 2 in GM for b=10 ms/μm$^2$. Diffusion times for all dMRI experiments were $\delta/\Delta$ =3/6 ms. The sample was kept at 37° C constantly during acquisition. Acquisition time was 2 hours for MGE and 53 hours for dMRI, where the sample should retain its tissue structure. No histology was performed after imaging.

*Data processing*

Data processing was done in Matlab (The MathWorks, Natick, MA, USA). All complex MRI images were denoised using tensor MP-PCA[50,51] with a window size of [7 7 7], and subsequently Gibbs-unrung[52] using the complex denoised images.

*MGE pipeline*

The complex signal phase was fitted to a linear function $\phi(t) = [\overline{\Omega}_{MRI} + \overline{\Omega}_{Bgf} + \overline{\Omega}_{Ref}]t + \phi_0$ based on the echo times above 20 ms, where $\phi_0$ accounts for unwanted $B_1$ effects. The frequency $\overline{\Omega}_{MRI} + \overline{\Omega}_{Bgf} + \overline{\Omega}_{Ref}$ was then unwrapped using SEGUE[53], and the Laplacian Boundary Value method[54] (LBV) was utilized for removing $\overline{\Omega}_{Bgf}$ +

$\overline{\Omega}_{\text{Ref}}$. Figure S2 in the supporting material gives an overview of the MGE pipeline showing both raw images, and the different processing steps for the phase.

*dMRI pipeline*

Figure S3 in the supporting material gives an overview of the dMRI pipeline. We averaged the dMRI across all echo times using singular value decomposition (SVD), to extract the diffusion-weighted signal component. After this we used the signal magnitude for fODF fitting. Due to sample drift between acquiring dMRI and MGE signals, a rigid co-registration of the dMRI signal to the MGE signal was necessary to align the fODF with the MGE signal.

*DKI and fODF fitting algorithms*

We estimated mean diffusivity (MD) and fractional anisotropy (FA) by fitting **exp1** data to the Diffusion Kurtosis Imaging[55,56] (DKI) signal expression. The fODF Laplace coefficients $p_{lm}$ were estimated from **exp2** data using Fiber Ball Imaging[23] (FBI), which is based on the "Standard Model" of diffusion in WM[39] (SM) and assumes the extra-axonal water signal is negligible for high gradients. We set the intra-axonal diffusivity to 2 μm²/ms. However, the effect of using a lower diffusivity on the fODF is small[23]. We used $l_{max} = 6$ for all methods.

*Susceptibility fitting algorithms*

Susceptibility fitting was done using an iterative least squares algorithm ([LSMR](.))[57]. When fitting ex vivo images, where no ground truth is available, we regularized the LSMR algorithms by selecting the number of iterations that maximized curvature of the L-curve[58,59], which depicts the trade-off between the least squares norm and the norm of the solution. Susceptibility was referenced to the PBS fluid in the lateral and third ventricles (see supplementary material S2 for more on referencing).

Three different frequency models were considered in this study:

- **MACRO** $\overline{\chi}_{\text{QSM}}$:

$$\text{argmin}_{\overline{\chi}_{QSM}} \left\| \overline{\Omega}_{\text{MRI}}(\mathbf{R}) - \gamma B_0 \mathbf{M}^{\text{Brain}}(\mathbf{R})\widehat{\mathbf{B}}^T \sum_{\mathbf{R}'} \mathbf{Y}(\mathbf{R} - \mathbf{R}')\overline{\chi}_{\text{QSM}}(\mathbf{R}')\widehat{\mathbf{B}} \right\|_2$$

(8)

$\overline{\chi}_{QSM}$ denotes the susceptibility fit without mesoscopic contribution (i.e. $\overline{\Omega}^{Meso}(\mathbf{R}) = 0$) and corresponds to standard QSM. Notice that we here used the elementary dipole field[10] $\mathbf{Y}(\mathbf{R} - \mathbf{R}')$ (no bars). $M^{Brain}(\mathbf{R})$ is the sample mask (not magnetization) enforcing the spatial distribution of measurements inside the brain[60].

- **MESO+MACRO $\overline{\chi}_{QSM+}$:**

$$\mathrm{argmin}_{\overline{\chi}_{QSM+}} \left\| \overline{\Omega}_{MRI}(\mathbf{R}) - \gamma B_0 M^{Brain}(\mathbf{R}) \left( -\frac{1}{3}\sum_{m=-2}^{2} p_{2m}(\mathbf{R}) Y_{2m}(\widehat{\mathbf{B}}) M^{WM}(\mathbf{R}) \overline{\chi}_{QSM+}(\mathbf{R}) + \widehat{\mathbf{B}}^T \sum_{\mathbf{R}'} \overline{\mathbf{Y}}(\mathbf{R} - \mathbf{R}') \overline{\chi}_{QSM+}(\mathbf{R}') \widehat{\mathbf{B}} \right) \right\|_2. \quad (9)$$

$\overline{\chi}_{QSM+}$ denotes susceptibility fit proposed here and includes mesoscopic contribution estimated using the $p_{2m}$ of the fODF, as well as the voxel-averaged dipole field $\overline{\mathbf{Y}}$. Here $M^{WM}(\mathbf{R})$ is a WM mask based on the fractional anisotropy of the scatter matrix generated from the fODF threshold at 0.45.

**STI $\overline{\chi}_{STI}$:**

$$\mathrm{argmin}_{\overline{\chi}_{STI}} \left\| \sum_{\widehat{\mathbf{B}}} \{\overline{\Omega}_{MRI}(\mathbf{R}; \widehat{\mathbf{B}}) - \gamma B_0 M^{Brain}(\mathbf{R}) \widehat{\mathbf{B}}^T \sum_{\mathbf{R}'} \mathbf{Y}(\mathbf{R} - \mathbf{R}') \overline{\chi}_{STI}(\mathbf{R}') \widehat{\mathbf{B}} \} \right\|_2. \quad (10)$$

$\overline{\chi}_{STI}$ denotes susceptibility fitting using STI. As for QSM, it is a purely macroscopic model, with the only difference being that now we fit a rank-2 susceptibility tensor using multiple sample (or $\widehat{\mathbf{B}}$) orientations.

*MRI experiment with multiple sample orientations*

In supplementary material S4 we have included an MRI experiment on an ex vivo rat brain at 9.4T. Here, MGE was acquired at 5 different sample orientations and dMRI at 1 orientation. Acquisition parameters are described in S4, with imaging and data processing similar to the mouse brain. Susceptibility fitting was done using Eqs. (8) and (9) for each sample orientation, and including all orientations at once corresponding to COSMOS[61] with and without incorporating mesoscopic frequency shifts.

**Digital brain phantom simulation**

We tested the accuracy in susceptibility fitting of the two models (**QSM** vs. **QSM+**) on a digital phantom (cf. Figure 3) with piece-wise constant susceptibility based on the FA and MD maps. The phantom includes both anisotropic myelin susceptibility and iron sources. We segmented the brain into WM and GM by creating a binary mask $M^{WM}(\mathbf{R})$ from high FA regions of the fODF scatter matrix threshold at 0.35. Notice this is lower than used in the fitting algorithm to emulate an unsuccessful segmentation of WM when fitting. From these, we synthesized 4 orientation invariant susceptibility parameters and computed their frequency contributions

$$\Delta\overline{\chi}(\mathbf{R}) = -1 \cdot FA(\mathbf{R}) \cdot M^{WM}(\mathbf{R}) \rightarrow \overline{\Omega}_{\Delta\chi}^{Meso}(\mathbf{R}) + \overline{\Omega}_{\Delta\chi}^{Macro}(\mathbf{R})$$

$$\overline{\chi}(\mathbf{R}) = 5 \cdot FA(\mathbf{R}) \cdot M^{WM}(\mathbf{R}) \rightarrow \overline{\Omega}_{\chi}^{Meso}(\mathbf{R}) + \overline{\Omega}_{\chi}^{Macro}(\mathbf{R})$$

$$\overline{\chi}_{WM}^{S}(\mathbf{R}) = \overline{\chi}_{WM}^{S} \cdot MD(\mathbf{R}) \cdot M^{WM}(\mathbf{R}) \rightarrow \overline{\Omega}_{\chi_{WM}^{S}}^{Macro}(\mathbf{R})$$

$$\overline{\chi}_{GM}^{S}(\mathbf{R}) = \overline{\chi}_{GM}^{S} \cdot MD(\mathbf{R}) \cdot \left(1 - M^{WM}(\mathbf{R})\right) \rightarrow \overline{\Omega}_{\chi_{GM}^{S}}^{Macro}(\mathbf{R}). \quad (11)$$

The sum of all frequencies defines the ground truth $\overline{\Omega}_{MRI}(\mathbf{R})$ of the phantom, assuming the reference frequency $\overline{\Omega}_{Ref}(\mathbf{R})$ has been removed and no background fields were present. The ratio $\overline{\chi}/\Delta\overline{\chi}$ between the two WM susceptibilities is based on previous findings[21], while $\overline{\chi}_{WM}^{S}$ and $\overline{\chi}_{GM}^{S}$ enable us to vary ratios of spherical susceptibility compared to WM. We assume mesoscopic contributions from spheres to be uniformly distributed in each voxel, so their mesoscopic contribution is zero. $\overline{\Omega}_{\chi_{GM}^{S}}^{Macro}(\mathbf{R})$ is computed like $\overline{\Omega}_{\chi}^{Macro}(\mathbf{R})$ in the second term of Equation (4).

Due to the absence of an analytical result for $\overline{\Omega}_{\Delta\chi}^{Meso}$, we simulated the Lorentzian tensor $\mathbf{L}_{\Delta\chi}$ for uniformly dispersed cylinders up to a cut-off angle $\theta_c$, as done in a similar manner in our previous study[22] (cylinder configurations can be seen in Figure S3 in supplementary material). Randomly positioned, non-overlapping single-layered cylinders, with a ratio between inner and outer radii of 0.6, are packed with a volume fraction of 15%. Their radii are varied following a gamma distribution (see Figure S4). To compute a mesoscopic contribution $\overline{\Omega}_{\Delta\chi}^{Meso}(\mathbf{R})$ in our phantom, we used the major fiber direction of the fODF along with its dispersion angle $\theta_{p2}$ [62] to define a new axially symmetric and cone shaped fODF with cut-off angle $\theta_{p2}$. We then used our simulation as a look-up table to estimate $\overline{\Omega}_{\Delta\chi}^{\text{Meso}}(\mathbf{R})$. To treat the $\overline{\Omega}_{\chi}^{\text{Meso}}(\mathbf{R})$ and $\overline{\Omega}_{\Delta\chi}^{\text{Meso}}(\mathbf{R})$ on equal footing, we used the same cone shaped fODF to compute their mesoscopic contributions.

Three phantoms of increasing complexity were investigated with different combinations of susceptibility. The three ground truths (GT) are shown in Figure 5 while the titles indicate the added sources. We generated the corresponding frequency shift for each phantom and added noise corresponding to an SNR=50. We then estimated the susceptibility using either Equations (8) or (9). We optimized the LSMR fitting algorithm for each GT and Equations (8) or (9) individually, by fitting with *l2 (Tikhonov)* regularization ranging from 1 to 0.002 in 50 logarithmically distributed steps. Through each iterative step in the LSMR algorithm, we computed the RMSE between our fitted susceptibility and the isotropic susceptibility sources of the ground truth, normalized to the norm of isotropic susceptibility sources[63]. The solution used for further analysis was then chosen based on the regularization and iteration step that minimized the RMSE. This was done to ensure a fair comparison with minimal bias caused by the ill-posed nature of the fitting problem. Upon fitting, the susceptibility maps were referenced to CSF, which we defined as having zero susceptibility.

*STI phantom*

We also synthesized an STI phantom including only WM for simplicity. We computed the Larmor frequency (including mesoscopic frequency contributions) at 21 unique sample orientations using electrostatic repulsion[64], both with and without susceptibility anisotropy, and then performed STI to estimate an apparent susceptibility tensor using Equation (10). We then compared the two cases in terms of their mean magnetic susceptibility MMS = $\frac{1}{3}(\chi_1 + \chi_2 + \chi_3)$, susceptibility anisotropy index MSI = $|\chi_1 - \chi_3|$ and color-coded MSI from the eigenvector of the eigenvalue closest to zero[17,28].

## Results

**Digital brain phantom**

Figure 4 shows the eigenvalues for the Lorentzian tensor $\mathbf{L}_{\Delta\chi}$ from susceptibility anisotropy and isotropic susceptibility used to compute the mesoscopic frequency shift. Figure 5A shows the resulting susceptibility fits for all three phantoms with different susceptibility sources along with the difference to ground truth. It is clear from the residuals that WM is less biased for QSM+ compared to QSM. Figure 5B shows the normalized RMSE for all three phantoms for different ratios of variances $\sigma^2\left(\overline{\chi}^S_{\text{WM(GM)}}(\mathbf{R})\right)/\sigma^2\left(\overline{\chi}(\mathbf{R})\right)$ between the spherical and cylindrical susceptibility in WM (variance within $M^{\text{WM}}$). Here we find that our constrained model has the lowest RMSE if $\sigma^2\left(\overline{\chi}(\mathbf{R})\right)$, associated with the bulk isotropic axonal susceptibility, is greater than $\sigma^2\left(\overline{\chi}^S_{\text{WM}}(\mathbf{R})\right)$ of the WM iron related susceptibility. The same was true when the ratio between the mean magnitude susceptibilities $\left\langle\left|\overline{\chi}^S_{\text{WM(GM)}}(\mathbf{R})\right|\right\rangle/\langle|\overline{\chi}(\mathbf{R})|\rangle$ was less than 1 (here $\langle\cdot\rangle$ denotes average across $M^{\text{WM}}$).

*STI brain phantom*

Figure 6 shows MMS, MSI and color-coded MSI for the phantom with and without WM susceptibility anisotropy $\Delta\overline{\chi}$. MMS and MSI only change 10% and 12% RMSE, when adding anisotropy. This shows that the mesoscopic contribution of WM fibers with susceptibility $\overline{\chi}$ are the main source of anisotropy and not actual susceptibility anisotropy.

**Ex vivo brain imaging**

*Magnetic susceptibility $\overline{\chi}$*

Figure 7 shows the susceptibility maps from two different coronal slices of the mouse brain (Sagittal and horizontal slices can be seen in Figure S5 and Figure S6, respectively). The last 2 rows show the susceptibility difference $\delta\bar{\chi}$ of $\overline{\chi}_{\text{QSM}}$ compared to $\overline{\chi}_{\text{QSM+}}$ with the fODF estimated from FBI at different *b*-values. We observed increased hyperintensity in highly anisotropic WM parallel to the main field such as the anterior commissure. Here we found a mean bulk WM susceptibility and standard deviation to be around -98±10 ppb (compared to -75±8 when mesoscopic contributions from WM are not included), which is closer to previous findings[15,21] than QSM.

*Larmor frequency contributions*

The macroscopic and mesoscopic contributions to the Larmor frequency were calculated using the forward relation in Equation (3) with the estimated susceptibility and $p_{2m}$ of the fODF as input. The result is shown in Figure 8. Figure 8A, clearly show that the mesoscopic contribution is non-zero in white-matter regions, and when the field is parallel to the axon, it is positive and opposite in sign to the macroscopic contribution, as expected from theory. Figure 8B shows a 3D maximum intensity projection of $\overline{\Omega}_{\text{MRI}}$, $\overline{\Omega}^{\text{Meso}}$ and $\overline{\Omega}^{\text{Macro}}$ (at b=10 ms/µm²). This demonstrates that $\overline{\Omega}^{\text{Meso}}$ provides a novel contrast by combining information of both $p_{2m}$ and $\overline{\chi}$.

*MRI experiment with multiple sample orientations*

In supplementary material S4, we show that the susceptibility obtained from COSMOS (cf. Figure S8) including mesoscopic frequency shifts produces slightly lower residuals with visually less structural bias in comparison to conventional COSMOS (cf. Figure S7). We also find that WM susceptibility becomes more negative by the mesoscopic correction, in agreement with the effect observed on the single orientation fit of the mouse brain (cf. Figure 7). For the single orientation susceptibility fits, we observed only a small improvement in the residuals (cf. Figure S9).

## Discussion

### Incorporation mesoscopic field effects into QSM

Estimating magnetic susceptibility is challenging for many reasons. In particular, the MRI measured Larmor frequency shift $\overline{\Omega}_{\text{MRI}}$ depends on the local organization of magnetized tissue at the mesoscopic scale. This contribution has so far not been included in standard quantitative susceptibility (QSM) models, but can potentially be responsible for a frequency shift on the same order of magnitude as the contribution from neighboring voxels, the only contribution considered in QSM[22]. In fact, this is why the average field outside long parallel randomly positioned cylinders in a cylindrical container is zero as the mesoscopic frequency shifts are equal to and opposite the macroscopic frequency contributions.

*Minimal magnetic microstructure model*

The purpose of this study was to develop a minimal model framework for the measured Larmor frequency when sampling at multiple orientations is not feasible. Our model includes frequency shifts from the white matter (WM) microstructure with microscopic isotropic susceptibility and structurally isotropic sources with isotropic susceptibility in gray matter. In reality, WM voxel contains multiple sources, for example highly aligned myelinated axons and non-heme iron[65]. However, our model offers an improvement in susceptibility estimation compared to QSM as long the variance or the mean magnitude in bulk susceptibility of e.g., WM iron is lower than for the isotropic bulk susceptibility of myelin - no matter if susceptibility anisotropy is present or not (see simulation in Figure 5).

*Future extensions of the biophysical model*

A reasonable next step will be to analytically include mesoscopic frequency contributions from microscopic WM susceptibility anisotropy to extend our model framework. Estimating model parameters for such a model requires sampling at multiple orientations. While our model includes myelin water (MW), evidence[45,66,67] suggests a large frequency shift in MW that goes beyond our proposed susceptibility model. For that reason, we assumed MW to be fully relaxed. This is a reasonable assumption since we only considered the signal phase at echo times above 20 ms, where MW should be absent at 16.4T due to its fast relaxation rate. Different mechanisms have been proposed to explain this observation[47,68,69], and we aim to investigate it in the future in order to include MW in our model. Additional frequency shifts from various randomly oriented magnetic inclusions with scalar susceptibilities, e.g., to model the effect of iron, will also be considered in the future. Modelling the signal relaxation within the same biophysical picture can also add additional information, which could be used to include e.g., an iron-related susceptibility without sample rotations[5].

**Limitations**

*Susceptibility and frequency contributions*

We estimated the bulk scalar magnetic susceptibility of an ex vivo mouse with and without including mesoscopic frequency contributions from WM (Figure 7). The susceptibility maps revealed noticeable differences in contrast and large quantitative differences. In the anterior commissure, the root-mean-squared-difference in $\bar{\chi}$ was 25%, when mesoscopic contributions from WM are not considered. A similar susceptibility difference was observed in an ex vivo rat brain (Figure S7-Figure S9 supplementary material S4) where we included multiple sample orientations in the susceptibility fit. This underscores the impact of including microstructural field effects when quantifying magnetic susceptibility, even without including tensor $\chi$. However, it is important to understand these mechanisms better in the future, before attempting to achieve robust susceptibility estimations and resolve multiple types of inclusions in a single voxel.

While our model only includes a single degree of freedom, we found that the ill-posed nature associated with the dipole field eroded the effect of the mesoscopic frequency shift. This was evident when comparing single orientation fits with a multi orientation fit (See Figure S7-Figure S9 in supplementary material S4). Here we found that the iterative LSMR algorithm used required many iterations (on the order of 100 iterations without any regularization due to having multiple orientations) in order for the residuals to be lower when incorporating the mesoscopic correction in WM. For the single orientation fits on the rat brain, we observed that the noise corrupted the fit after around 5-10 iterations, when no regularization was included. When including an *l2* Tikhonov regularization, a higher number of iterations could be reached, but at the expense of a larger bias in susceptibility values and in the residuals in Larmor frequency - especially in WM where the susceptibility was highest, ultimately eroding the improvement by the mesoscopic correction. Hence, while our model only includes one degree of freedom, it still benefits from acquiring images at multiple sample orientations to make the inverse problem better posed, or by using better fitting algorithms with more sophisticated regularization schemes than Tikhonov regularization.

Nevertheless, while our WM model is simple compared to actual magnetic tissue microstructure, we believe the model's apparent susceptibility gives an important first insight into the relationship between mesoscopic and macroscopic frequency contributions in real data.

*fODF*

The fiber orientation distribution (fODF) was estimated by doing spherical decomposition of the dMRI signal at high b using Fiber Ball Imaging[23] (FBI). As a flavor of the Standard Model of diffusion in white matter[39] (SM), it models WM axons similarly to our proposed WM axon model. In comparison to DTI-derived metrics, such as FA and the primary diffusion eigenvector which describe the diffusive dispersion from both intra- and extra-axonal diffusion anisotropy, SM-derived methods estimating the fODF allows estimating the actual fiber orientation dispersion.

SM considers dispersion between bundles of parallel axons, while our susceptibility model considers dispersion between individual fibers. Nevertheless, our model is consistent with the axon configuration in SM, since bundles of randomly positioned parallel cylinders does not give rise to any additional frequency shift[22].

Even though misestimation of the fODF will bias susceptibility estimates, only the *l=2* expansion coefficients, $p_{2m}$, of the fODF are necessary to estimate the mesoscopic frequency shifts. These are typically rather robust against noise, and with less variation across different diffusion times[70].

It took around 53 hours to acquire dMRI signals used for fODF estimation. While this is far beyond a reasonable timeframe in a clinical setting, a normal FBI protocol could be done in around 10 minutes on a clinical scanner [71]. The large scan time here was chosen to achieve ultra-high isotropic resolution (100μm isotropic) with high SNR, to reduce image artifacts and achieve optimal co-registration between dMRI and MGE voxels. For this we used a 3D acquisition with no partial Fourier acceleration or acceleration scheme such as EPI.

*Fixation effects*

As imaging was performed on ex vivo mouse brains, effects related to fixation may also affect the estimated parameters due to structural alterations, increased chemical shifts and changes in chemical composition[15,20,49,72]. Susceptibility values have earlier been found to be numerically smaller in vivo compared to ex vivo[15,73]. For example, phosphate-buffered saline (PBS) and paraformaldehyde (PFA) solution can lead to increased macro-molecular exchange, earlier found to lead to shifts on the order of -0.013ppm and 0.05ppm respectively[15]. Secondly, PFA susceptibility differs by -0.028ppm compared to CSF[15].

**Implications for QSM and STI**

So far, QSM has been regarded as the best option for susceptibility estimation, when rotating the sample is not possible. Our simulations indicate that the best strategy for the simplest possible susceptibility model is to include only the largest contributor to the Larmor frequency in each voxel. In WM, this is believed to be the isotropic component of the myelin susceptibility tensor[21,43]. Equation 3 represents the Larmor frequency shift in our model framework including mesoscopic frequency shifts from WM microstructure. As it is seen from our ex vivo fitting, including mesoscopic frequency shifts in WM can substantially change susceptibility estimation. This requires estimating the fODF at high b-value, optimally around $b$=10 ms/µm$^2$.

Susceptibility tensor imaging (STI) represents a natural extension of QSM to include macroscopic tensor anisotropy while still neglecting mesoscopic frequency shifts. Numerous studies have applied the STI model as a demonstration of WM susceptibility anisotropy[17,28,29]. However, microstructurally related frequency shifts in WM produce a large bias in STI[21]. This was corroborated in a recent work[29] incorporating orientation dependent WM frequency offsets in STI fitting, resulting in a large decrease in susceptibility anisotropy on human brain. However, the susceptibility and fODF dependence in these local frequency offsets, which was demonstrated here (Figure 4) and in previous work[22], was not included.

Our simulations reveal that a predominant source of anisotropy in the STI tensor arises instead from the mesoscopic frequency from WM microstructure with only scalar susceptibility, i.e., microstructural anisotropy. In fact, the apparent anisotropy was the same order of magnitude as the mean susceptibility, and in line with experimental findings for STI[17,28,29]. We also compared our maps to known STI tractography studies[17,28,29], and found results strikingly similar to previous studies - including their characteristic deviation from standard DTI tractography. Second, when we include actual susceptibility anisotropy, we found that this only changed the measured STI tensor around 10% root-mean-squared-difference, indicating that a large sources of anisotropy in STI may originate from a mesoscopic contribution of WM *microstructure,* and not magnetic susceptibility anisotropy.

## Conclusion

We developed a novel minimal framework for including mesoscopic Larmor frequency contributions in quantitative susceptibility mapping (QSM), especially relevant when imaging at multiple orientations is not an option. This was done by modelling the frequency induced from white matter (WM) magnetic microstructure as organized in long multi-layered cylinders with orientation dispersion and scalar susceptibility. Through computer simulations, we find that our model improves susceptibility estimation compared to QSM, and Susceptibility Tensor Imaging (STI) are substantially biased by the unaccounted-for structural anisotropy due to the mesoscopic frequency contribution, indicating the observed STI tensor might not represent susceptibility anisotropy as expected. Our experimental results show that local WM microstructure induce a substantial frequency shift in WM and should not be ignored in QSM. We believe our results will advance the pursuit of a full characterization of magnetic microstructure of nervous tissue, with the goal of faithful parameter estimations that can be used actively in clinical research.

## Acknowledgments

This study is funded by the Independent Research Fund (grant 8020-00158B). We also thank Shemesh lab members, especially PhD Cristina Chavarrías and MSc Beatriz Cardoso for assisting with MRI experiments, and Prof. Mark D Does and Dr. Kevin Harkins from Vanderbilt University for the REMMI pulse sequence.

## Abbreviations

 **QSM**: Quantitative Susceptibility Mapping. **GE**: Gradient-recalled Echo Signal. **WM**: White Matter. **STI**: Susceptibility Tensor Imaging**. fODF**: Fiber Orientation Distribution Function. **FBI**: Fiber Ball Imaging. **dMRI**: Diffusion MRI. **GM**: Gray Matter. **STF**: Symmetric Trace Free Tensor. **SO(3)**: Group of all Rotations about the Origin in Three Dimensional Euclidian Space $\mathbb{R}^3$. **PBS:** Phosphate-buffered Saline. **PFA:** paraformaldehyde. **MGE**: Multi Gradient Echo-Recalled Signal. **MP-PCA**: Marchenko-Pastur Principal Component Analyses. **SVD:** Singular Value Decomposition. **MD**: Mean Diffusivity. **FA**: Fractional Anisotropy of diffusion tensor. **DKI**:

Diffusion Kurtosis Imaging. **SM**: Standard Model of Diffusion. **LSMR**: Least Squares Minimal Residual Method. **CSF**: Cerebral Spinal Fluid. **GM**: Gray Matter. **RMSE**: Root-mean-squared error.

# Figures

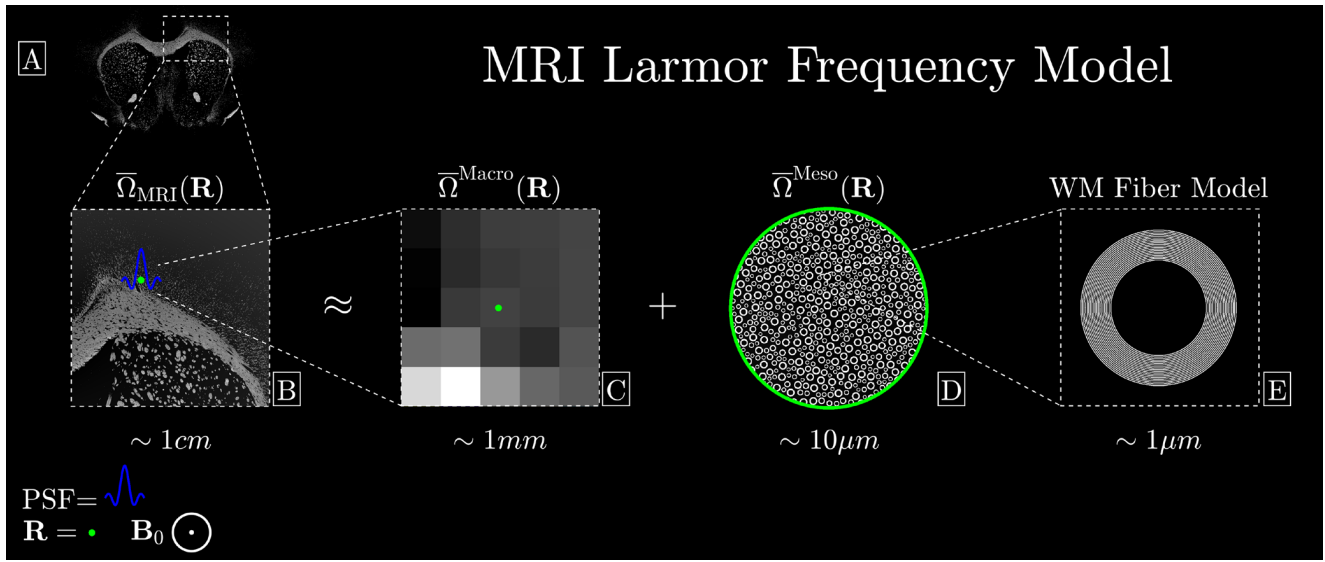

Figure 1

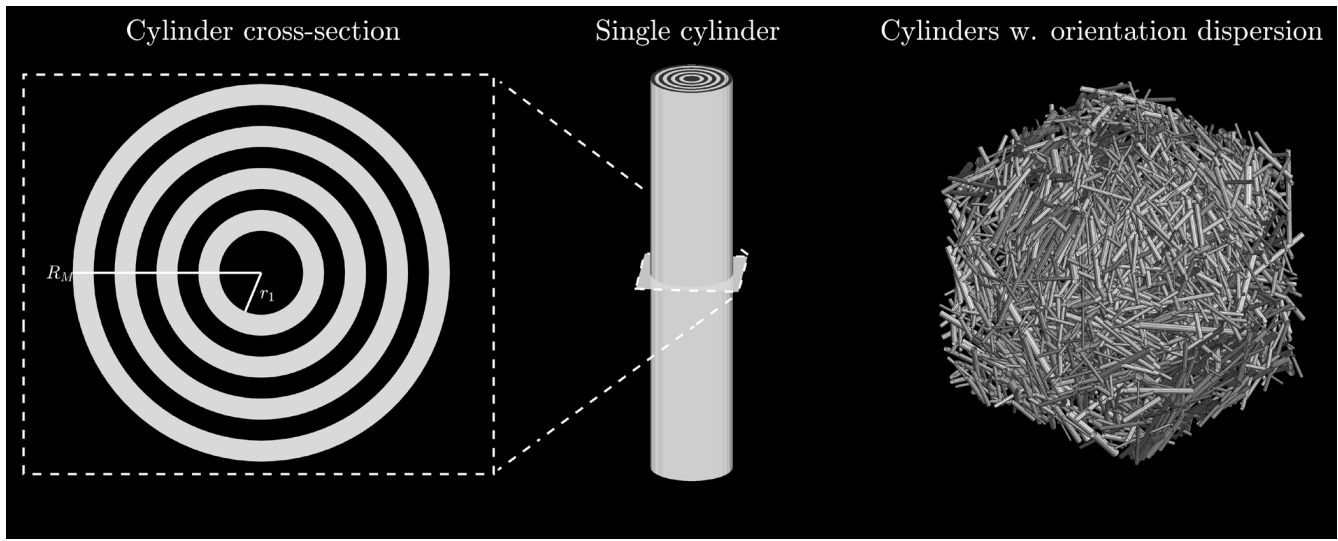

Figure 2

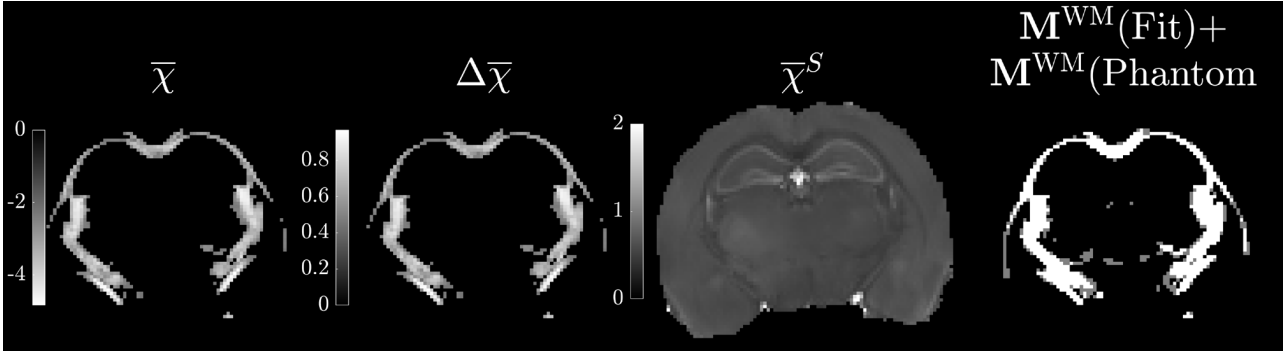

Figure 3

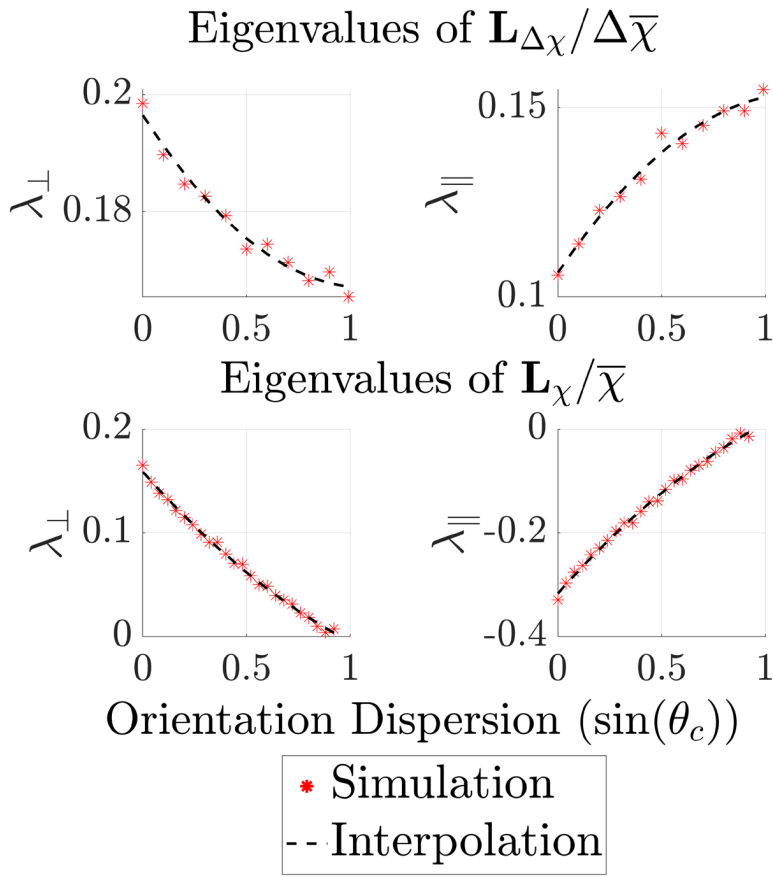

Figure 4

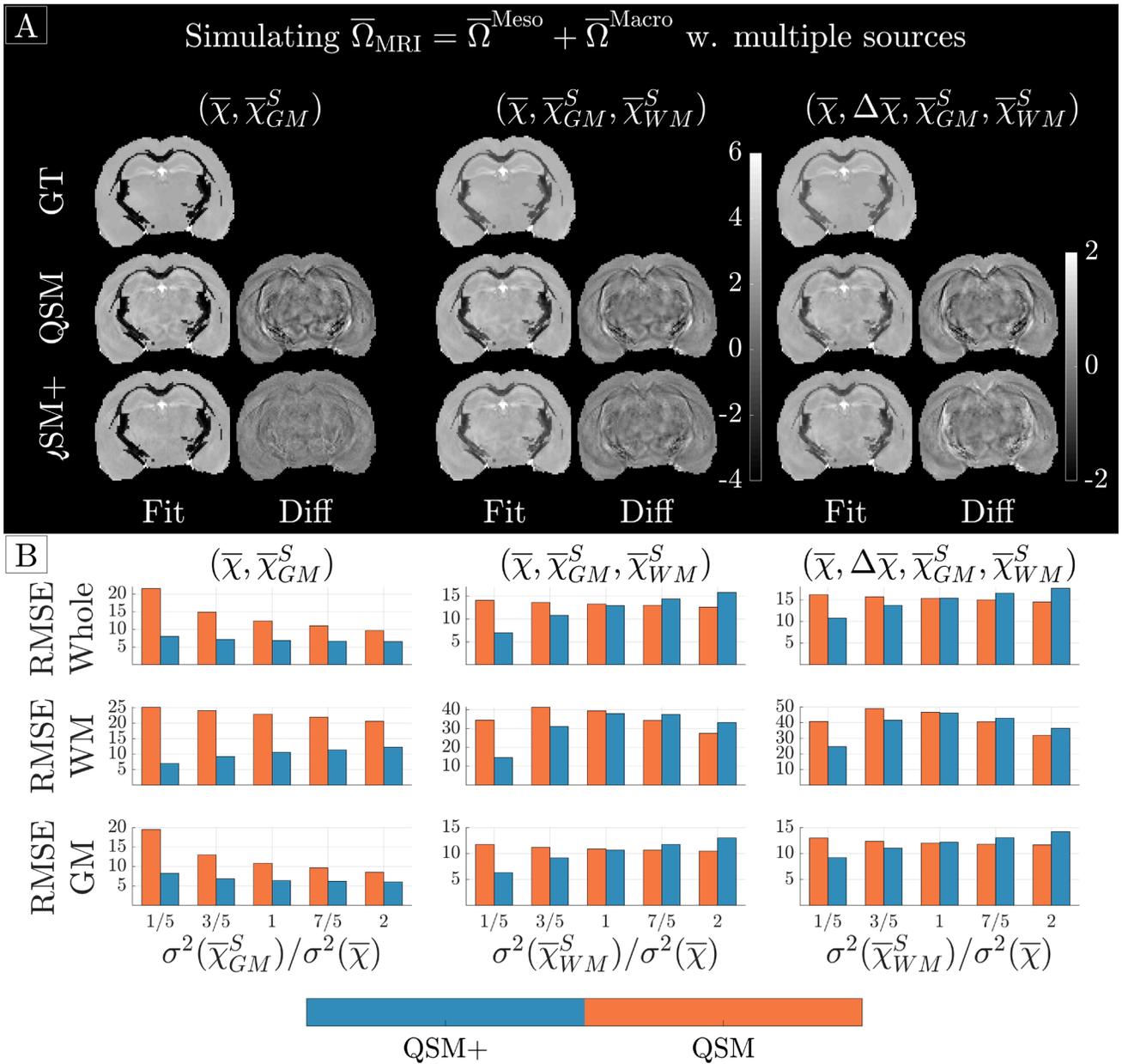

Figure 5

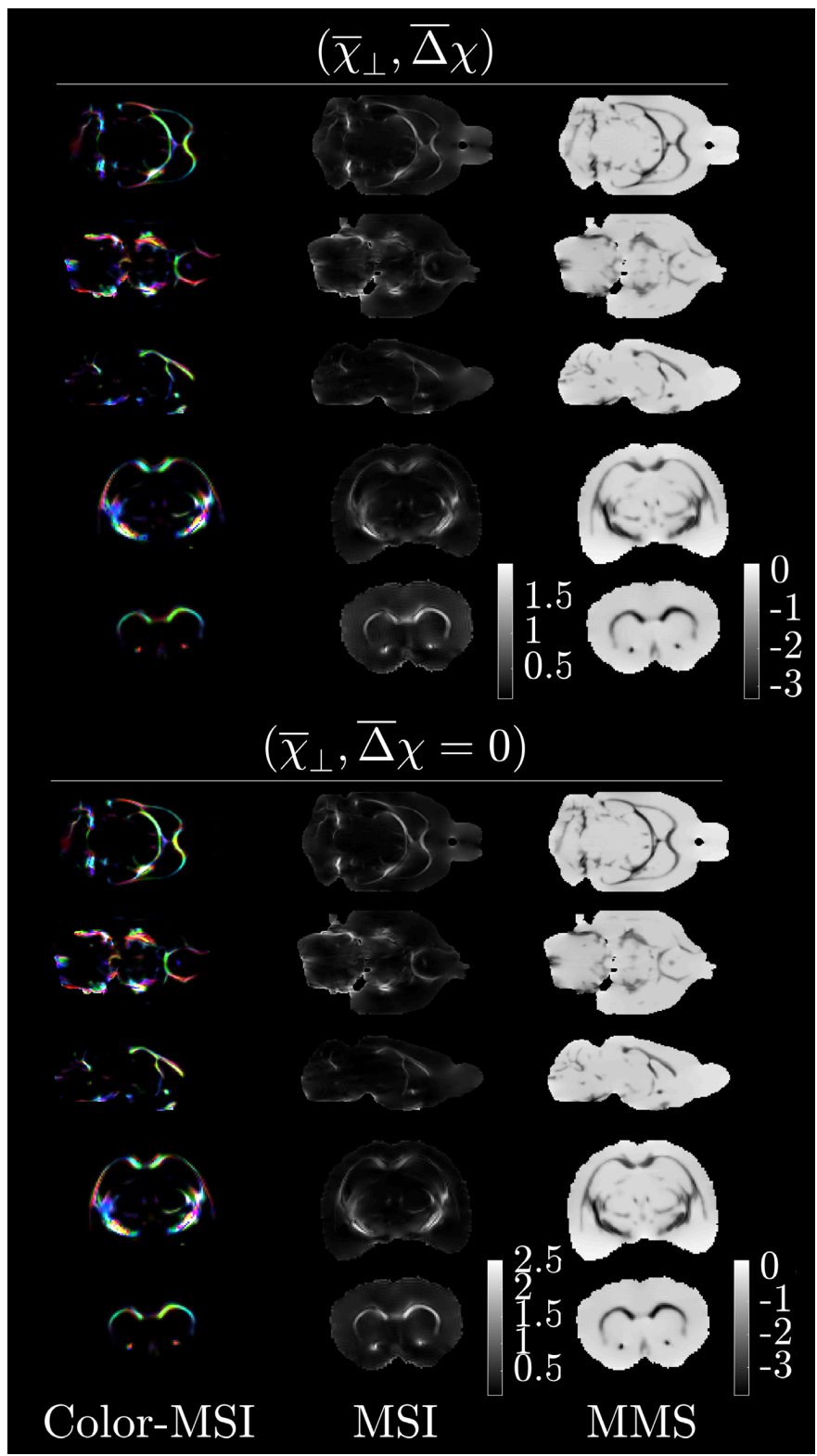

Figure 6

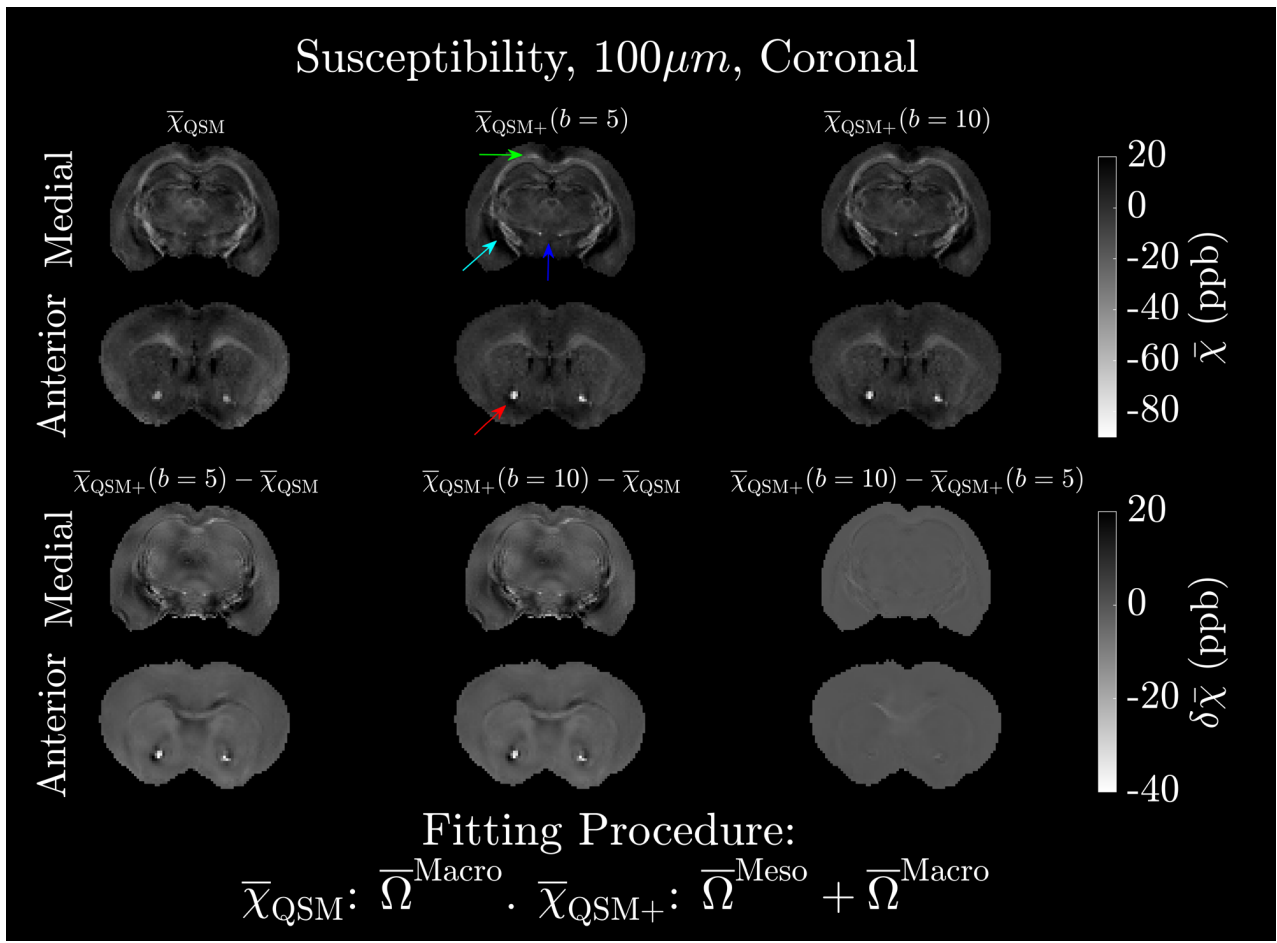

Figure 7

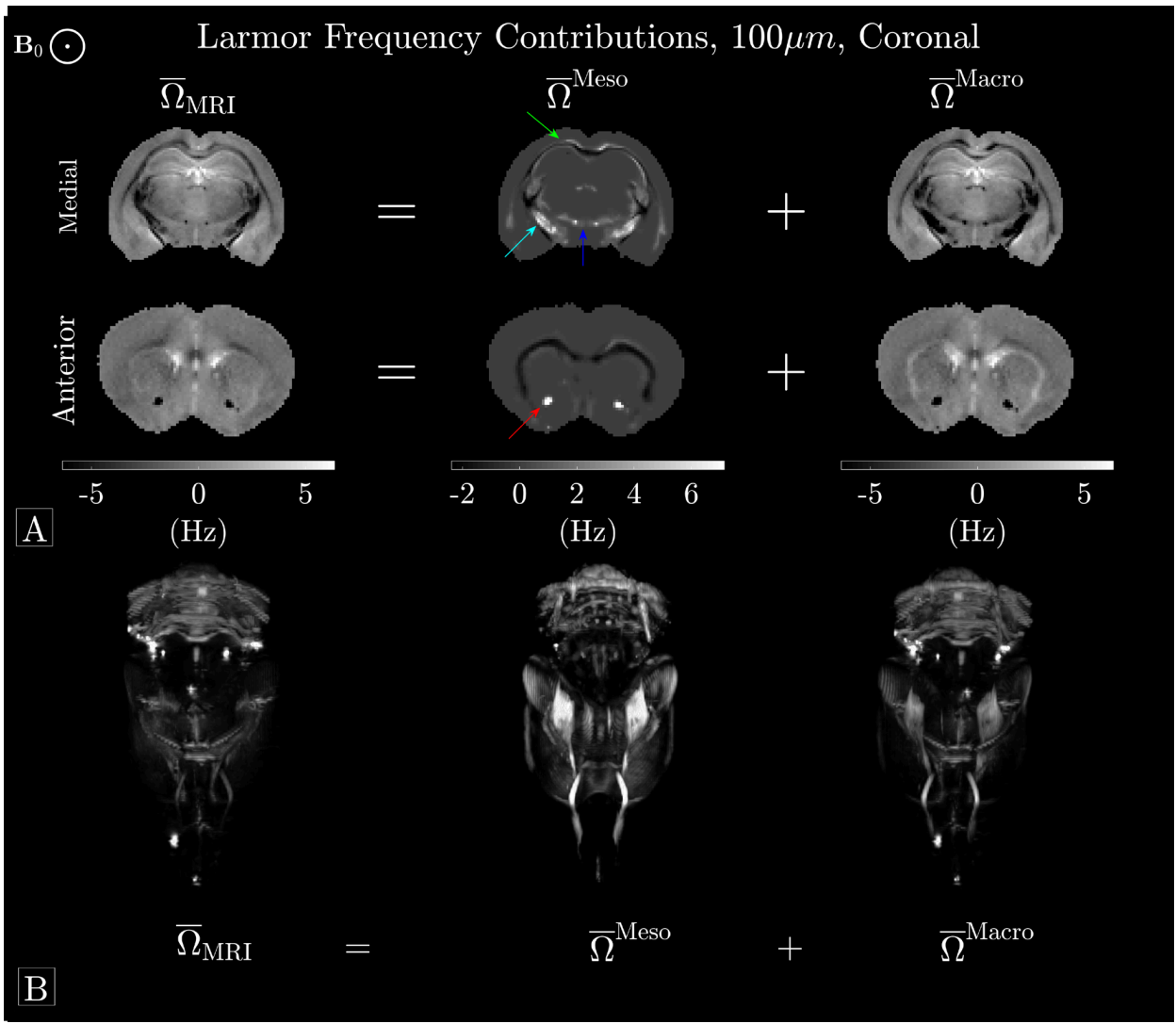

Figure 8

# Figure Captions

**Figure 1 - Model of the MRI Larmor frequency:** **A**: Myelin-stained coronal slice of mouse brain. **B**: The MRI measured Larmor frequency $\overline{\Omega}_{\text{MRI}}(\mathbf{R})$, coarse grained on the mesoscopic scale and sampled at discrete points **R**. Sampling is described by the point-spread-function (PSF), here shown as a blue sinc- function, whose width is macroscopic. For a slowly varying magnetic microstructure, $\overline{\Omega}_{\text{MRI}}(\mathbf{R})$ can be approximated by the following two contributions: **C**: The macroscopic contribution approximated at the scale of the sampling resolution capturing contributions at macroscopic distances; and **D**, the contribution from nearby magnetic microstructure within a mesoscopic Lorentz sphere. The latter contains here randomly placed multi-layered cylinders, one of which is depicted in E. Mouse brain image is reproduced from © 2011 Allen Institute for Brain Science, Allen Mouse Brain Connectivity Atlas, https://connectivity.brain-map.org.

**Figure 2 - Structural model of the WM mesoscopic environment:** Each fiber is modelled as M concentric cylinders of radii $r_j$ to $R_j$ (small/capital letters indicate inner/outer radii) with $j = 1, \ldots, M$. The cross-sectional volume fraction of the m'th fiber is $\zeta_m = \pi \sum_j (R_j^2 - r_j^2)$. The mesoscopic environment consists of N fibers with overall cross-sectional volume fraction $\zeta = \sum_m \zeta_m$ and a given orientation dispersion assumed to be independent of fiber positions and radii. Cylinders are impermeable with water uniformly distributed intra- and extra-cylindrical, and in between bilayers.

**Figure 3 - Susceptibility phantom**: Synthesized magnetic susceptibility of WM and spheres (iron). WM mask $\text{M}^{\text{WM}}$(phantom) is generated from a high FA mask with a threshold of 0.4. For fitting we used an FA mask $\text{M}^{\text{WM}}$ with a threshold of 0.45 to emulate an unsuccessful estimation of the total mesoscopic contribution. This is here demonstrated by their sum to show their overlap.

**Figure 4 - Simulation of the mesoscopic contribution from different orientation distributions:** Eigenvalues $(\lambda_\perp, \lambda_\parallel)$ of the Lorentzian tensor from WM susceptibility $\mathbf{L}_{\overline{\chi}}/\overline{\chi}$ and susceptibility anisotropy $\mathbf{L}_{\Delta\overline{\chi}}/\Delta\overline{\chi}$ are presented for various levels of dispersion set by the maximum allowed polar angle $\theta_c$. $\mathbf{L}_{\Delta\overline{\chi}}/\Delta\overline{\chi}$ was simulated for 12 different dispersions, while $\mathbf{L}_{\overline{\chi}}/\overline{\chi}$ is reproduced from previous study[22]. The black line shows the interpolation of the data to a second order polynomial, which was used as a look-up table for computing the mesoscopic frequency shifts from different fiber directions. The depicted perpendicular eigenvalue is the mean of the two perpendicular eigenvalues.

**Figure 5 - Fitting with and without adding mesoscopic frequency shifts:** Three different phantoms considered with different susceptibility contributions (as shown in titles). $\overline{\chi}$ is the WM axon susceptibility, while $\overline{\chi}_{WM}^S$ and $\overline{\chi}_{GM}^S$ is the spherical susceptibility in WM and GM, respectively. **A**: The first row shows the ground truth susceptibility for each of the three phantoms. Middle row shows fitting without adding mesoscopic contributions (QSM), while the bottom row shows fitting with (QSM+). Differences from ground truth are shown in the adjoining columns. SNR here is 50 with a 3/5 ratio between $\sigma^2(\overline{\chi})$ and $\sigma^2(\overline{\chi}_{WM}^S)$. **B** shows bar plot of normalized (compared to isotropic susceptibility of ground truth) RMSE referenced to CSF. The x-axis shows various ratios in variance where 1/5 to 3/5 are WM plausible, while the remaining are GM/thalamus plausible. When $\sigma^2(\overline{\chi})$ is greater or comparable with $\sigma^2(\overline{\chi}_{WM}^S)$, the lowest RMSE is achieved including mesoscopic frequency shifts to the model, even though it neglects the spatial heterogeneity of $\overline{\chi}_{WM}^S$ and WM susceptibility anisotropy.

**Figure 6 - Tensor eigenvalues of STI phantom:** Fitting results from applying STI to the measured Larmor frequency (Equation (10)) of a digital phantom sampled at 21 orientations, including both mesoscopic and macroscopic frequency shifts. The upper panel shows fitting the phantom with susceptibility anisotropy, while the bottom panel has no susceptibility anisotropy. Both phantoms include structural anisotropy due to the mesoscopic contribution. Comparing the mean magnetic susceptibility (MMS), anisotropy (MSI) and color-coded MSI using the eigenvector of the most positive susceptibility eigenvalue, we find that the source of anisotropy and tractography contrast stems not from susceptibility anisotropy, but rather a bias from pure structural anisotropy from the mesoscopic contribution with scalar susceptibility $\overline{\chi}$.

**Figure 7 - Susceptibility maps of mouse brain at 100 μm isotropic resolution:** Coronal slices from the medial and anterior parts of the brain are shown. $\overline{\chi}_{QSM}$ corresponds to zero mesoscopic contribution (analogous to QSM), and $\overline{\chi}_{QSM+}$ corresponds to a non-zero mesoscopic contribution calculated using this method. Largest differences are visible near the cingulum and corpus callosum (green), cerebral peduncle (light blue), and anterior commissure olfactory limb (red) and mammalithalamic tract (dark blue).

**Figure 8 - Macroscopic and mesoscopic Larmor frequencies (using $\overline{\chi}_{QSM+}$ at b = 10 ms/μm²):** **A**: The frequencies are calculated using the forward relation in Equation (3). The biggest mesoscopic contributions to the Larmor frequency are found in regions of highly anisotropic WM. This is especially visible near the cingulum and corpus callosum (green), cerebral peduncle (light blue), and anterior commissure olfactory limb (red) and mammalithalamic tract (dark blue). **B**: Horizontal 3D rendition of mesoscopic frequency $\overline{\Omega}^{Meso}$ at b=10ms/ μm² based on the maximum intensity projection.

**Figure S1 - Susceptibility phantom WM iron and myelin including external fields:** **A** shows the ground truth susceptibility $\delta\overline{\chi}_{GT}(\mathbf{R})$ from WM iron, WM myelin, GM iron, while the ventricles have zero susceptibility. **B** shows the corresponding frequency shift $\overline{\Omega}_{MRI}(\mathbf{R})$ from $\delta\overline{\chi}_{GT}(\mathbf{R})$ including frequency shifts from a uniform external and internal susceptibility. **C** shows $\overline{\Omega}_{MRI}(\mathbf{R})$ after background field removal, which removes both contributions from the internal and external uniform susceptibility. **D** Shows the difference between the fitted susceptibility $\delta\overline{\chi}_{Fit}(\mathbf{R})$ (after referencing to CSF in ventricles) and ground truth susceptibility $\delta\overline{\chi}_{GT}(\mathbf{R})$.

**Figure S2 - Overview of pipeline for MGE processing:** All the complex MGE images were MP-PCA denoised, and Gibbs-unrung. The complex phase was extracted, unwrapped and background-field corrected, and subsequently fitted to extract $\overline{\Omega}_{MRI}$. $\overline{\Omega}_{Background}$ shows the subtracted background frequency. Representative signal magnitude (left plot) and unwrapped and background-field corrected phase (right plot) are plotted for a white matter (cingulum in blue) and gray matter (thalamus in orange) voxel, respectively. Magnitude is shown in semi-log scale to illustrate the mono-exponential behavior of both signals are predominantly mono exponential. The phase behaves linearly in both WM and GM.

**Figure S3 - Overview of dMRI pipeline for data processing:** The Complex dMRI images were tensor MP-PCA denoised for each echo time individually followed by Gibbs-unringing. The signal magnitudes were then averaged over echo times using SVD, and the resulting images were then fitted with DKI or FBI for tensor or fODF estimation. Color-coded FA maps from diffusion tensor ($FA_D$) and scatter matrices ($FA_T$, cf. Equation (12) in supplementary material S2) from FBI are shown for various protocols. $S(b, \hat{\mathbf{g}})$ denotes the dMRI signal with b-value along $\hat{\mathbf{g}}$, here the in-plane direction $\hat{\mathbf{z}}$ (green on sphere).

**Figure S4 - Populations of cylinders with different levels of orientation dispersion are shown in A. B** shows the probability density function (pdf) of the resulting cylinder parameters for each configuration. The cylinder radius $\rho$ is gamma-distributed, while $\theta$ and $\varphi$ are uniformly distribution in the full range of azimuthal angle and from zero to the maximum polar angle $\theta_c$, respectively. Colors are used to represent different populations with orientation dispersion indicated by the colorbar.

**Figure S5 - Susceptibility maps of mouse brain at 100 μm isotropic resolution:** Horizontal slices from the medial and anterior parts of the brain are shown. $\overline{\chi}_{QSM}$ corresponds to zero mesoscopic contribution (analogous to QSM), and $\overline{\chi}_{QSM+}$ corresponds to a non-zero mesoscopic contribution calculated using this method.

**Figure S6 - Susceptibility maps of mouse brain at 100 μm isotropic resolution:** Sagittal slices from the medial and anterior parts of the brain are shown. $\overline{\chi}_{QSM}$ corresponds to zero mesoscopic contribution (analogous to QSM), and $\overline{\chi}_{QSM+}$ corresponds to a non-zero mesoscopic contribution calculated using this method.

**Figure S7 – COSMOS Susceptibility fitting of rat brain at 150 μm isotropic resolution:** The plot to the left show voxel-by-voxel comparison of the residuals $\delta\overline{\Omega}_{MRI}$ for fitting including all orientations. The red line corresponds to the unit line, while the blue shows a linear fit, with slope below 1, indicating lower residuals with QSM+. $\sigma_{\overline{B}}^2(\delta\overline{\Omega}_{MRI})$ shows the variance in the residuals for a coronal slice of the rat brain in the anterior part of the brain.

**Figure S8– COSMOS Susceptibility maps of rat brain at 150 μm isotropic resolution:** Coronal slices from the anterior part of the brain are shown. $\bar{\chi}_{QSM}$ corresponds to zero mesoscopic contribution (conventional COSMOS), and $\bar{\chi}_{QSM+}$ includes a non-zero mesoscopic contribution calculated using this method.

**Figure S9 - Susceptibility fitting of rat brain at 150 μm isotropic resolution at 5 different orientations:** The plots to the left show voxel-by-voxel comparison of the residuals $\delta\overline{\Omega}_{MRI}$ for each sample orientation labeled in the title. Nan corresponds to no rotation (two individual experiments are shown), and here the field is along the sagittal orientation of the brain. The red line corresponds to the unit line, while the blue shows a linear fit, with slope slightly below 1, indicating lower residuals with QSM+. $\sigma_{\overline{B}}^2(\delta\overline{\Omega}_{MRI})$ and $\sigma_{\overline{B}}^2(\delta\overline{\chi})$ show the variance in the residuals and susceptibility fits, respectively, for a coronal slice of the rat brain in the anterior part of the brain.

# Incorporating the effect of white matter microstructure in the estimation of magnetic susceptibility in ex-vivo mouse brain

## S1 Larmor frequency from WM with axons and uniform iron content

In this section we investigate the Larmor frequency caused by the magnetic susceptibility of the sample when white matter (WM) contains both myelinated axons and iron. We assume that iron content varies very little across WM. We consider here for simplicity a porous media of WM, gray matter (GM) and Cerebral-spinal fluid (W). We describe the macroscopic tissue regions by the non-overlapping indicator functions $1 = M^{WM}(\mathbf{R}) + M^{GM}(\mathbf{R}) + M^{W}(\mathbf{R})$ for $\mathbf{R}$ inside the brain. The WM compartment includes myelinated axons with susceptibility $\chi^C$. Iron complexes with $\chi^S$ are found in both GM and WM, while water with $\chi^W$ are in both WM, GM and W. The microscopic susceptibility thus becomes $\chi(\mathbf{r}) = v^C(\mathbf{r})\chi^C + v^S(\mathbf{r})\chi^S + v^W(\mathbf{r})\chi^W$ where $v$ are non-overlapping microscopic indicator functions fulfilling $v^C(\mathbf{r}) + v^S(\mathbf{r}) + v^W(\mathbf{r}) = 1$. Here we neglected susceptibility anisotropy of myelin for simplicity. The microscopic Larmor frequency offset thus becomes

$$\Omega(\mathbf{r}) = \gamma B_0 \hat{\mathbf{B}}^T \int d\mathbf{r} \left( v^C(\mathbf{r})\chi^C + v^S(\mathbf{r})\chi^S + v^W(\mathbf{r})\chi^W \right) \Upsilon(\mathbf{r} - \mathbf{r}') \hat{\mathbf{B}}.$$

(S13)

As a first step we rewrite Equation (S1) such that the tissue susceptibilities are referenced to the susceptibility $\chi^W$

$$\Omega(\mathbf{r}) = \gamma B_0 \hat{\mathbf{B}}^T \int d\mathbf{r} \cdot \left( v^C(\mathbf{r}')\delta\chi^C + v^S(\mathbf{r}')\delta\chi^S + \chi^W \right) \Upsilon(\mathbf{r} - \mathbf{r}') \hat{\mathbf{B}},$$

(S14)

where $\delta\chi = \chi - \chi^W$. To describe the measured frequency shift, we consider the mesoscopically averaged frequency shift $\overline{\Omega}(\mathbf{R})$ of Equation (S2) in terms of its mesoscopic and macroscopic contributions

$$\overline{\Omega}(\mathbf{R}) = \overline{\Omega}_{\delta\chi^S}^{\text{Meso}}(\mathbf{R}) + \overline{\Omega}_{\delta\chi^C}^{\text{Meso}}(\mathbf{R}) + \overline{\Omega}_{\delta\chi^S}^{\text{Macro}}(\mathbf{R}) + \overline{\Omega}_{\delta\chi^C}^{\text{Macro}}(\mathbf{R}) + \overline{\Omega}^{W}(\mathbf{R}).$$

(S15)

We assume that iron is homogenously distributed in each voxel in both WM and GM such that $\overline{\Omega}_{\delta\chi^S}^{\text{Meso}}(\mathbf{R}) = 0$. The iron WM susceptibility is further rewritten as the deviation from the mean $\delta\overline{\chi}^S$ across $M^{WM}$, i.e., $M^{WM}(\mathbf{R})\delta\chi^S(\mathbf{R}) = M^{WM}(\mathbf{R})\left(\delta\overline{\chi}^S + \left(\delta\chi^S(\mathbf{R}) - \delta\overline{\chi}^S\right)\right)$. If the variation in the mesoscopically averaged bulk susceptibility $\delta\chi^S(\mathbf{R}) - \delta\overline{\chi}^S$ in WM is sufficiently small compared to $\delta\overline{\chi}^C(\mathbf{R})$ from myelin, such that we may neglect it as a first order approximation, similar to why we neglected susceptibility anisotropy (Wharton & Bowtell, 2015) (see simulation in Figure 5 in the main text), then the macroscopic contribution from iron becomes

$$\overline{\Omega}_{\delta\chi^S}^{\text{Macro}}(\mathbf{R}) \approx \gamma B_0 \widehat{\mathbf{B}}^T \sum_{\mathbf{R}'} \overline{\mathbf{Y}}(\mathbf{R} - \mathbf{R}') \left(\delta\overline{\chi}^S M^{WM}(\mathbf{R}') + \delta\overline{\chi}^S(\mathbf{R}') M^{GM}(\mathbf{R}')\right) \widehat{\mathbf{B}}$$

$$= \gamma B_0 \widehat{\mathbf{B}}^T \sum_{\mathbf{R}'} \overline{\mathbf{Y}}(\mathbf{R} - \mathbf{R}') \left(\left(\delta\overline{\chi}^S(\mathbf{R}') - \delta\overline{\chi}^S\right) M^{GM}(\mathbf{R}') + \delta\overline{\chi}^S\left(1 - M^W(\mathbf{R}')\right)\right) \widehat{\mathbf{B}}.$$

(S16)

Using Equations (S3), (S4) and (4), the mesoscopically averaged frequency shift becomes

$$\overline{\Omega}(\mathbf{R}) = -\gamma B_0 \delta\overline{\chi}^C(\mathbf{R}) \frac{1}{3} \sum_{m=-2}^{2} p_{2m}(\mathbf{R}) Y_{2m}(\widehat{\mathbf{B}}) + \gamma B_0 \widehat{\mathbf{B}}^T \sum_{\mathbf{R}'} \overline{\mathbf{Y}}(\mathbf{R} - \mathbf{R}') \left(\left(\delta\overline{\chi}^S(\mathbf{R}') - \delta\overline{\chi}^S\right) M^{GM}(\mathbf{R}') - \delta\overline{\chi}^S M^W(\mathbf{R}') + \chi^W + \delta\overline{\chi}^S\right) \widehat{\mathbf{B}}.$$

(S17)

The macroscopic contribution of the constant susceptibility $\chi^W + \delta\overline{\chi}^S$ defines the reference frequency $\overline{\Omega}_{\text{Ref}}(\mathbf{R})$ in Equation (4) and is removed by the background field removal algorithm (Schweser et al., 2017). We see that when referencing to water after susceptibility fitting, WM is described by both myelin and WM iron through $\delta\overline{\chi}^C(\mathbf{R}) + \delta\overline{\chi}^S$, while GM describes its iron content $\delta\overline{\chi}^S(\mathbf{R})$, all in reference to water.

This is demonstrated in Figure S1A depicting a simple brain phantom with uniform susceptibilities $\delta\overline{\chi}^C = -5$, $M^{WM}(\mathbf{R})\delta\overline{\chi}^S = 4$, $M^{WM}(\mathbf{R})\delta\overline{\chi}^S = 5$. Susceptibility in ventricles is set to zero as we referenced to water $\delta\chi^W = 0$. This defines the ground truth susceptibility $\delta\overline{\chi}_{GT}(\mathbf{R})$, which we wish to estimate. We also get a susceptibility component $\overline{\chi}^W = -9$ across the whole sample after referencing to water and we also include a uniform external susceptibility $\overline{\chi}^{Ext} = 20$. The Larmor frequency $\overline{\Omega}_{MRI}(\mathbf{R})$ was computed using Equation (4) with $\gamma B_0 = 1$ including mesoscopic frequency shifts from myelin. We computed $\overline{\Omega}_{MRI}(\mathbf{R})$ from 10 unique sample orientations made using electrostatic repulsion(Jones et al., 1999) to avoid magic-angle artifacts. Figure S1B shows the corresponding Larmor frequency including mesoscopic frequency shifts. We removed the reference frequency $\overline{\Omega}_{Ref}(\mathbf{R})$ and external field contribution using LBV (Zhou et al., 2014), as is seen in Figure S1C. We then estimated the susceptibility $\delta\overline{\chi}_{Fit}(\mathbf{R})$ using Equation (9) and referenced it to the mean susceptibility in the ventricles. This referencing removes any constant susceptibility component across the sample caused by LBV when removing external fields. Figure S1D shows the difference $\delta\overline{\chi}_{Fit}(\mathbf{R}) - \delta\overline{\chi}_{GT}(\mathbf{R})$ in susceptibility to ground truth $\delta\overline{\chi}_{GT}(\mathbf{R})$. In WM, we find a mean susceptibility -0.98±0.35 in agreement with $\delta\overline{\chi}^C + M^{WM}(\mathbf{R})\delta\overline{\chi}^S$, in GM we find 4.93±0.41 corresponding to $M^{GM}(\mathbf{R})\delta\overline{\chi}^S$ and in CSF 0±0.14 corresponding to $\delta\overline{\chi}^W$.

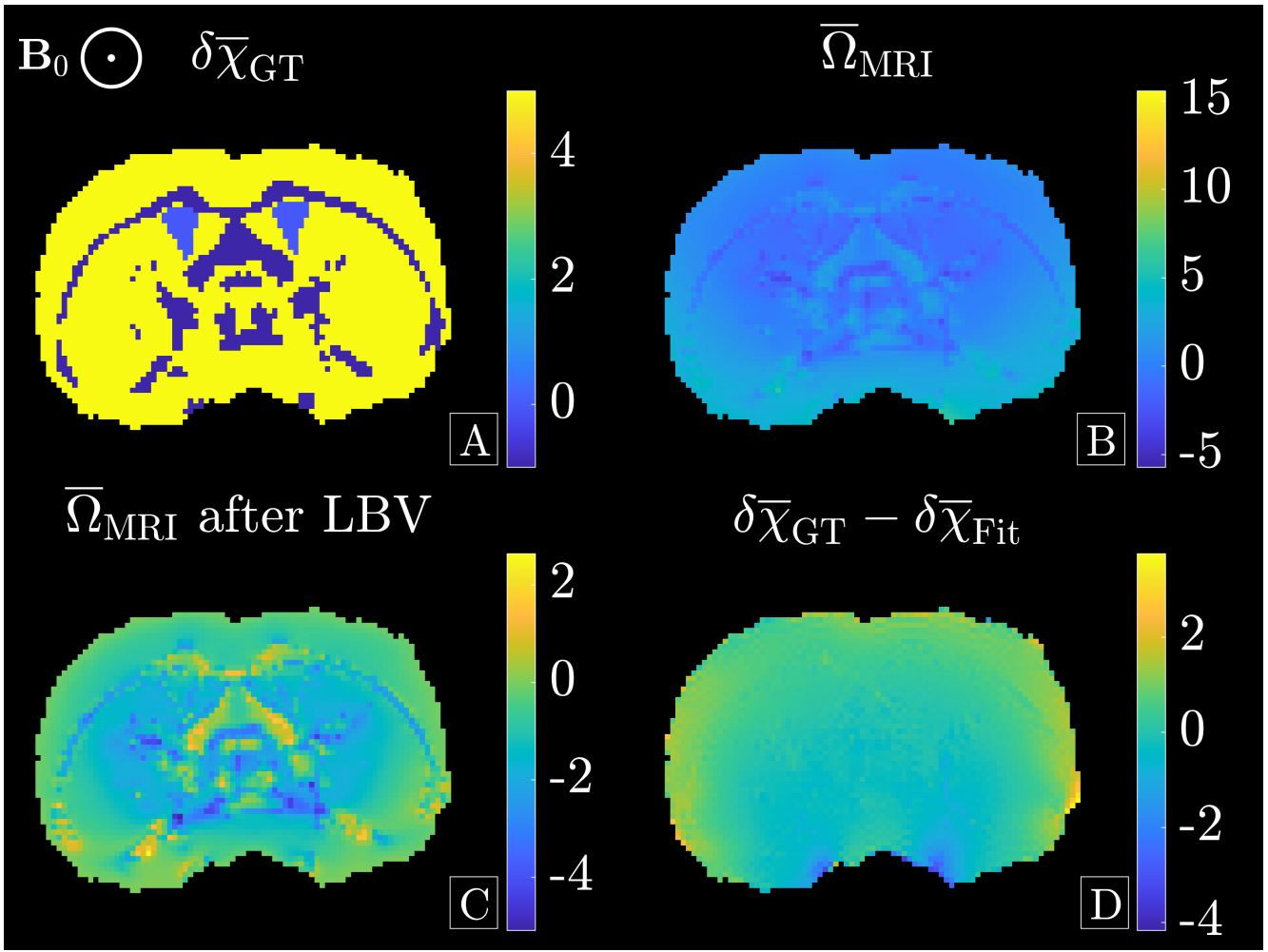

**Figure S10 - Susceptibility phantom WM iron and myelin including external fields:** **A** shows the ground truth susceptibility $\delta\overline{\chi}_{GT}(\mathbf{R})$ from WM iron, WM myelin, GM iron, while the ventricles have zero susceptibility. **B** shows the corresponding frequency shift $\overline{\Omega}_{MRI}(\mathbf{R})$ from $\delta\overline{\chi}_{GT}(\mathbf{R})$ including frequency shifts from a uniform external and internal susceptibility. **C** shows $\overline{\Omega}_{MRI}(\mathbf{R})$ after background field removal, which removes both contributions from the internal and external uniform susceptibility. **D** Shows the difference between the fitted susceptibility $\delta\overline{\chi}_{Fit}(\mathbf{R})$ (after referencing to CSF in ventricles) and ground truth susceptibility $\delta\overline{\chi}_{GT}(\mathbf{R})$.

## S2   Indicator function for multilayer cylinder

In this section we derive the mesoscopic demagnetization tensor (Sandgaard, Shemesh, et al., 2022), **N**, for multi-layer cylinders with arbitrary orientations (see Figure 2) to extend our model for solid cylinders. As described in previous work, the mesoscopic demagnetization tensor (Sandgaard, Shemesh, et al., 2022) **N** depends only on structural correlations,

$$\mathbf{N} = \frac{1}{(1-\overline{\zeta})} \int \frac{d\mathbf{k}}{(2\pi)^3} \mathbf{\Upsilon}(\mathbf{k}) \Gamma^{vv}(\mathbf{k}). \tag{S18}$$

Here $\Gamma^{vv}$ is the structural correlation function, whose generic form in Fourier space is

$$\Gamma^{vv}(\mathbf{k}) = \frac{v(\mathbf{k})v(-\mathbf{k})}{|\mathrm{M}|}, k > 0, \tag{S19}$$

and zero for $k = 0$. When susceptibility is uniform, the product $\mathbf{L} = -\chi \mathbf{N}$ defines the mesoscopic Lorentzian tensor (Kiselev, 2019; Sandgaard, Shemesh, et al., 2022) and characterizes $\overline{\Omega}^{\mathrm{Meso}}(\mathbf{R})$ (cf. Equation (2)). The indicator function $v(\mathbf{k})$ for an infinitely long cylinder consisting of $M$ concentric shells is a superposition of $2M$ solid infinite cylinders(Sandgaard, Kiselev, et al., 2022)

$$v(\mathbf{k}) = e^{iu\mathbf{k}\cdot\hat{\mathbf{u}}} \frac{4\pi^2}{k} \sum_{q=1}^{M} \left( R_q J_1(R_q k) - r_q J_1(r_q k) \right) \delta(\mathbf{k} \cdot \hat{\mathbf{n}})$$

$$= 2\pi e^{iu\mathbf{k}\cdot\hat{\mathbf{u}}} v^{2D}(k) \delta(\mathbf{k} \cdot \hat{\mathbf{n}}). \quad \text{(Multi layer cylinder )}$$

$$\tag{S20}$$

Here $v^{2D}(k)$ defines the indicator function in the 2D plane transverse to the orientation $\hat{\mathbf{n}}$, where $R_q, r_q$ denotes the outer and inner radii of the $q$'th layer. Consider N multilayer cylinders as conceptualized in Figure 2. They are randomly positioned and exhibit arbitrary orientation dispersion independent of their size. Summing over all N multilayer cylinders the total correlation function $\Gamma^{vv}(\mathbf{k})$, Equation (S7), splits into a sum over autocorrelation $\Gamma^{\mathrm{Auto}}(\mathbf{k})$ and cross-correlation $\Gamma^{\mathrm{Cross}}(\mathbf{k})$

$$\Gamma^{vv}(\boldsymbol{k}) = \Gamma^{\text{Auto}}(\boldsymbol{k}) + \Gamma^{\text{Cross}}(\boldsymbol{k}), \tag{S21}$$

where

$$\Gamma^{\text{Auto}}(\boldsymbol{k}) = \sum_m \Gamma_m(\boldsymbol{k}) \tag{S22}$$

and

$$\Gamma^{\text{Cross}}(\boldsymbol{k}) = \sum_{m \neq w} \Gamma_{mw}(\boldsymbol{k}). \tag{S23}$$

The total mesoscopic demagnetization tensor **N** relates to each of the two correlation functions by the sum **N** = **N**$^{\text{Auto}}$ + **N**$^{\text{Cross}}$, where each contribution is computed like in Equation (S6) using either Equation (S10) or (S11). Using Equations (S7) and (S8), we find for $\Gamma_m(\boldsymbol{k})$ in Equation (S10)

$$\Gamma_m(\boldsymbol{k}) = \Gamma_m^{2D}(k)\delta(\boldsymbol{k} \cdot \hat{\boldsymbol{n}}_m). \tag{S24}$$

The form of the autocorrelation $\Gamma_m(\boldsymbol{k})$ is identical to that of solid cylinders considered previously (Sandgaard, Shemesh, et al., 2022), i.e., it described by a 2D correlation function $\Gamma_m^{2D}(k)$ in the plane perpendicular to $\hat{\boldsymbol{n}}_m$. Using Equation (S10), the contribution from autocorrelations **N**$^{\text{Auto}}$,

$$\mathbf{N}^{\text{Auto}} = \frac{1}{(1-\zeta)} \sum_m \int \frac{d\boldsymbol{k}}{(2\pi)^3} \boldsymbol{\Upsilon}(\boldsymbol{k}) \Gamma^{\text{Auto}}(\boldsymbol{k}), \tag{S25}$$

is given be the radial and angular integrals (Sandgaard, Shemesh, et al., 2022), respectively

$$\frac{1}{(1-\zeta)} \int \frac{dk\, k}{(2\pi)^2} \Gamma_m^{2D}(k) = \frac{1}{(1-\zeta)} \Gamma_m^{2D}(r=0) = \zeta_m \tag{S26}$$

$$\int \frac{d\hat{\boldsymbol{k}}}{2\pi} \boldsymbol{\Upsilon}(\hat{\boldsymbol{k}}) \delta(\hat{\boldsymbol{k}} \cdot \hat{\boldsymbol{n}}_m) = \left(\frac{1}{3}\mathbf{I} - \frac{1}{2}(\mathbf{I} - \hat{\boldsymbol{n}}_m \hat{\boldsymbol{n}}_m^{\mathbf{T}})\right). \tag{S27}$$

We thus obtain for the autocorrelation contribution **N**$^{\text{Auto}}$

$$\begin{aligned} \mathbf{N}^{\text{Auto}} &= \sum_m \zeta_m \left( \tfrac{1}{3}\mathbf{I} - \tfrac{1}{2}(\mathbf{I} - \hat{\mathbf{n}}_m \hat{\mathbf{n}}_m^{\text{T}}) \right) \\ &= \zeta \left( \mathbf{T} - \tfrac{1}{6}\mathbf{I} \right) \\ &= \zeta \tfrac{1}{3} \sum_{m=-2}^{2} p_{2m} \mathbf{\mathcal{Y}}_{2m}. \end{aligned} \quad (S28)$$

Here $\mathbf{T} = \langle \hat{\mathbf{n}}^T \hat{\mathbf{n}} \rangle$ is the scatter matrix (Fisher et al., 1987), which was rewritten in terms of $p_{2m}$, the Laplace expansion coefficients of the fODF. $\mathbf{\mathcal{Y}}_{2m}$ is the symmetric trace-free tensors (STF) corresponding to an irreducible rank-2 representation of SO(3) (Thorne, 1980). The cross correlation $\Gamma_{mw}(\mathbf{k})$, Equation (S11), corresponds to a sum of $4M$ cross-correlations from solid cylinders, which we previously found not to contribute (Sandgaard, Shemesh, et al., 2022). We can thus set $\Gamma_{mw}(\mathbf{k}) = 0$ resulting in $\mathbf{N}^{\text{Cross}} = 0$. This then yields the same mesoscopic dipole tensor as for solid cylinders

$$\mathbf{N} = \mathbf{N}^{\text{Auto}} = \zeta \tfrac{1}{3} \sum_{m=-2}^{2} p_{2m} \mathbf{\mathcal{Y}}_{2m}. \quad (S29)$$

**A1)** *Compartmental average Larmor frequency $\overline{\Omega}_C^{\text{Meso}}$*

Here we briefly outline why Equation (S17) also corresponds to the mesoscopic dipole tensor in each of the three major water compartments. This means that the mesoscopic contribution to the average field in the extra-cylindrical compartment is the same as the intra-cylindrical compartment, and across bi-layers. Instead of relating the water indicator function directly to the negated indicator function of the cylinder, we may also characterize each major water compartment by their total indicator functions $v_I(\mathbf{k})$, $v_B(\mathbf{k})$ and $v_E(\mathbf{k})$, respectively

$$v_I(\mathbf{k}) = \sum_m e^{i\mathbf{k}\cdot\mathbf{u}_m} \tfrac{4\pi^2}{k} r_1 J_1(kr_1) \delta(\mathbf{k} \cdot \hat{\mathbf{n}}_m), \text{ (intra-cylindrical)}$$

$$v_B(\mathbf{k}) = \sum_m e^{i u_m \mathbf{k}\cdot\mathbf{u}_m} \tfrac{4\pi^2}{k} \sum_{q=2}^{M} \left( R_{(q-1)} J_1(r_{(q-1)}k) - r_q J_1(r_q k) \right) \delta(\mathbf{k} \cdot \hat{\mathbf{n}}_m), \text{ (bi-layers)}$$

$$v_E(\mathbf{k}) = (2\pi)^3 \delta(\mathbf{k}) - v_I(\mathbf{k}) - v_B(\mathbf{k}) - v(\mathbf{k}), \text{ (extra-cylindrical)} \quad (S30)$$

Hence, the structural correlation function is $v(\mathbf{k}) = (2\pi)^3 \delta(\mathbf{k}) - v_I(\mathbf{k}) - v_B(\mathbf{k}) - v_E(\mathbf{k})$. From this we can define the mesoscopic contribution to the compartmental Larmor frequency $\overline{\Omega}_C^{\text{Meso}}$:

$$\overline{\Omega}_C^{\text{Meso}} = -\gamma B_0 \chi \hat{\mathbf{B}}^T \mathbf{N}_C^{\text{Meso}} \hat{\mathbf{B}}, \tag{S31}$$

where the compartmental mesoscopic demagnetization tensor $\mathbf{N}_C^{\text{Meso}}$ depends on the compartmental correlation functions $\Gamma_C(\mathbf{k})$

$$\mathbf{N}_C^{\text{Meso}} = -\frac{1}{\zeta_C} \int \frac{d\mathbf{k}}{(2\pi)^3} \mathbf{Y}(\mathbf{k}) \Gamma_C(\mathbf{k}), \qquad \Gamma_C(\mathbf{k}) = \frac{v_C(\mathbf{k}) v(\mathbf{k})}{|M|} - \zeta_C \zeta (2\pi)^3 \delta(\mathbf{k}). \tag{S32}$$

$\Gamma_C(\mathbf{k})$ is a cross-correlation as it describes correlations between the water compartment defined by $v_C(\mathbf{k})$ with volume fraction $\zeta_C$ and the microstructure with indicator function $v(\mathbf{k})$, Equation (S8), with volume fraction $\zeta$. Using Equations (S14)-(S15) in Equation (S20), and that $\Gamma_C(r=0) = \zeta_C \zeta$, yields identical mesoscopic dipole tensors for all compartments:

$$\mathbf{N}_C^{\text{Meso}} = \zeta \frac{1}{3} \sum_{m=-2}^{2} p_{2m} \mathbf{\mathcal{Y}}_{2m}. \tag{S33}$$

Thus every compartment experiences the same average magnetic field, and the weighted sum $\mathbf{N}^{\text{Meso}} = \sum_C \frac{\zeta_C}{1-\zeta} \mathbf{N}_C^{\text{Meso}} = \mathbf{N}_C^{\text{Meso}}$ corresponds to Equation (S17) as expected. This means that if we filter the signal through diffusion weighting to isolate intra-cylindrical signals, we do not gain any new information about the magnetic microstructure.

## S3) Supplementary Figures

Here we present supplementary figures for the article.

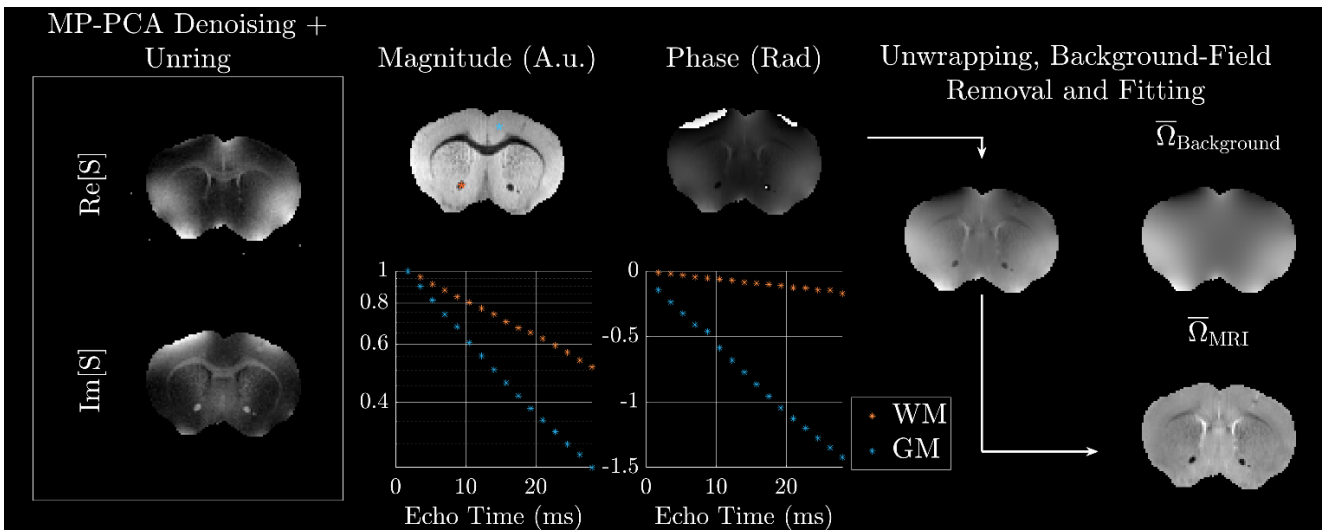

**Figure S2 - Overview of pipeline for MGE processing:** All the complex MGE images were denoised using MP-PCA followed by Gibbs-unringing. The complex phase was extracted, unwrapped and background-field corrected, and subsequently fitted to extract $\overline{\Omega}_{MRI}$. $\overline{\Omega}_{Bgf}$ shows the subtracted background frequency, using a *depth* and *peel* set to 3 to erode field errors from fluid accumulated on the surface of brain. Representative signal magnitude (left plot) and unwrapped and background-field corrected phase (right plot) are plotted for a white matter (cingulum in blue) and gray matter (thalamus in orange) voxel, respectively. Magnitude is shown in semi-log scale to illustrate the mono-exponential behavior of both signals are predominantly mono exponential. The phase behaves linearly in both WM and GM.

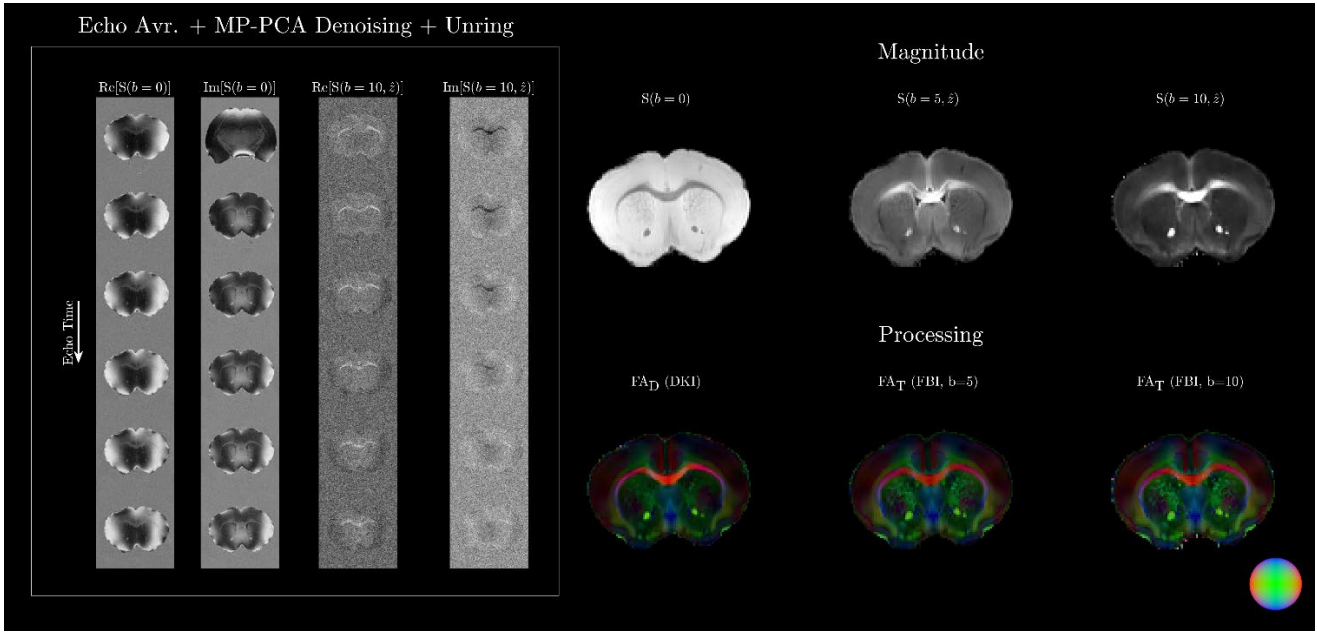

**Figure S3 - Overview of dMRI pipeline for data processing:** The Complex dMRI images were tensor MP-PCA denoised for each echo time individually followed by Gibbs-unringing. The signal magnitudes were then averaged over echo times using SVD, and the resulting images were then fitted with DKI or FBI for tensor or fODF estimation. Color-coded FA maps from diffusion tensor ($FA_D$) and scatter matrices ($FA_T$, cf. Equation (S34) in appendix A) from FBI are shown for various protocols. $S(b, \hat{\mathbf{g}})$ denotes the dMRI signal with b-value along $\hat{\mathbf{g}}$, here the in-plane direction $\hat{\mathbf{z}}$ (green on sphere).

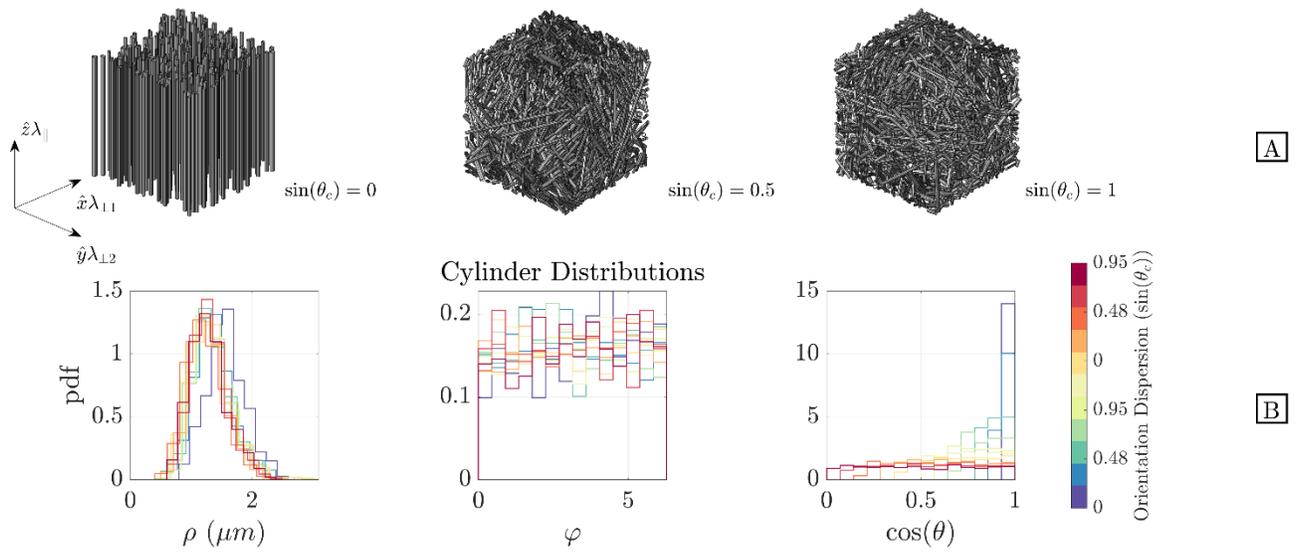

**Figure S4 - Populations of cylinders with different levels of orientation dispersion are shown in A. B** shows the probability density function (pdf) of the resulting cylinder parameters for each configuration. The cylinder radius $\rho$ is gamma-distributed, while $\theta$ and $\varphi$ are uniformly distribution in the full range of azimuthal angle and from zero to the maximum polar angle $\theta_c$, respectively. Colors are used to represent different populations with orientation dispersion indicated by the colorbar.

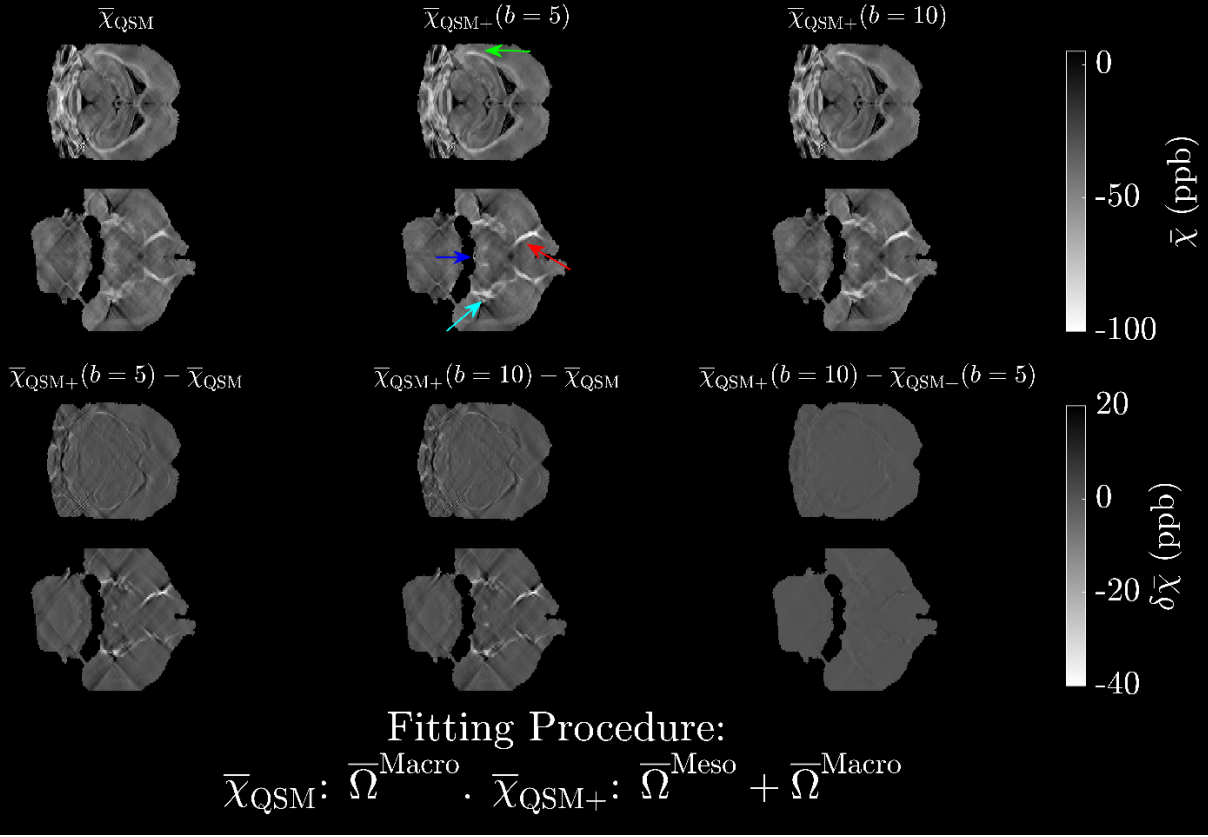

**Figure S5 - Susceptibility maps of mouse brain at 100 μm isotropic resolution:** Horizontal slices from the medial and anterior parts of the brain are shown. $\bar{\chi}_{QSM}$ corresponds to zero mesoscopic contribution (analogous to QSM), and $\bar{\chi}_{QSM+}$ corresponds to a non-zero mesoscopic contribution calculated using this method.

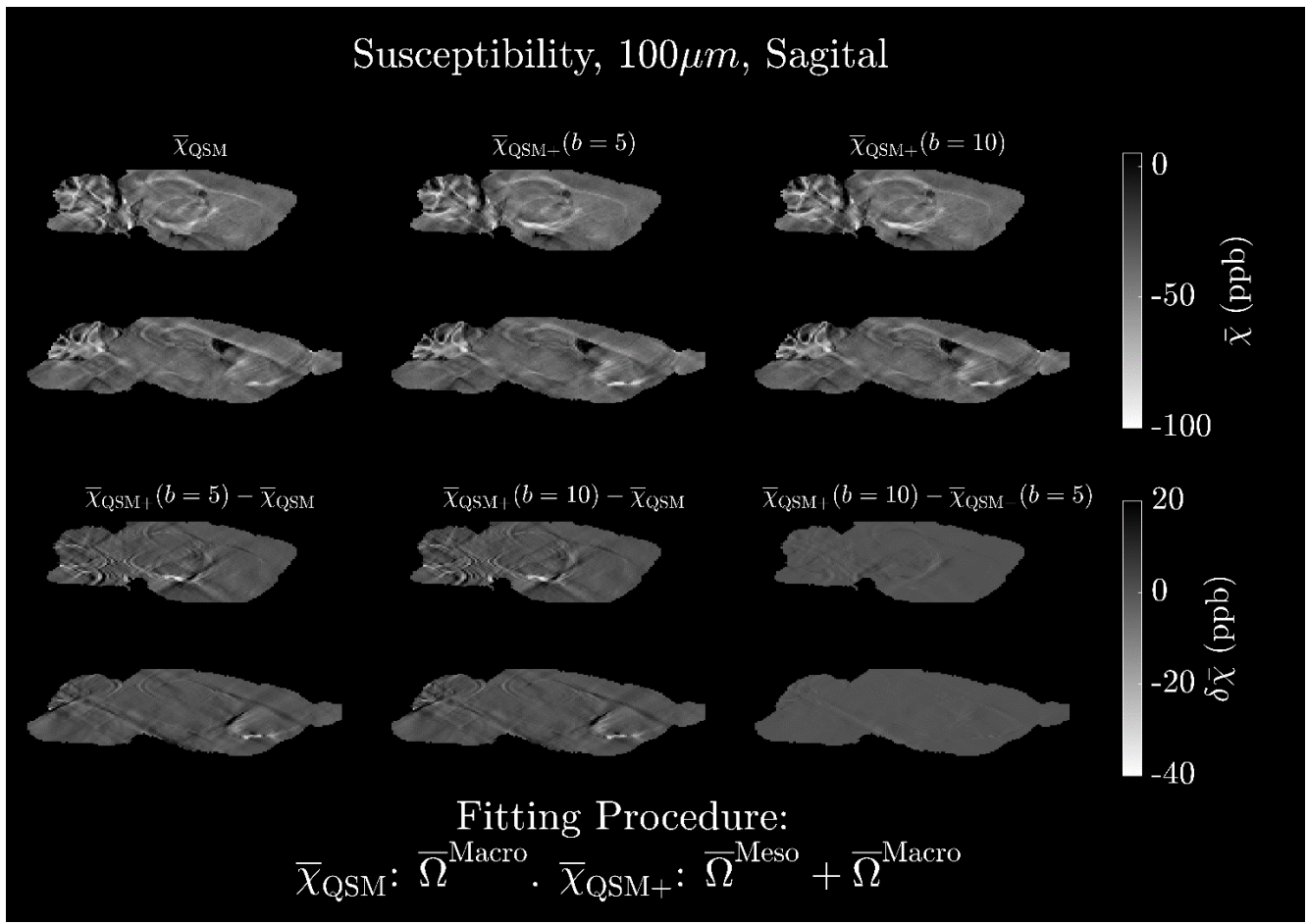

**Figure S6 - Susceptibility maps of mouse brain at 100 μm isotropic resolution:** Sagittal slices from the medial and anterior parts of the brain are shown. $\bar{\chi}_{\text{QSM}}$ corresponds to zero mesoscopic contribution (analogous to QSM), and $\bar{\chi}_{\text{QSM+}}$ corresponds to a non-zero mesoscopic contribution calculated using this method.

# S4) Susceptibility fitting with a single orientation versus multiple orientations in an ex vivo rat brain

In this supplementary section we investigate the QSM quality from single orientation susceptibility fitting compared to a COSMOS fit (Liu et al., 2009), which means multiple directions are included in the QSM fit to overdetermine the inverse problem. We consider fitting without and with the addition of the mesoscopic frequency shift $\overline{\Omega}^{Meso}$ described by Equations (8) (QSM) and (9) (QSM+) in the main text, respectively.

Here we demonstrate that COSMOS including a mesoscopic frequency shift in WM offers the lowest residuals $\delta\overline{\Omega}_{MRI}$ between the measured Larmor frequency and the predicted from fitting. This is also the case for single orientation fitting, but the improvement is very small due to the fitting algorithm being too sensitive to noise after only a few iterations.

## Methods

*Ex vivo brain imaging*

All animal experiments were preapproved by the competent institutional and national authorities and carried out according to European Directive 2010/63.

*Animal preparation*

The Animal experiment were performed on a perfusion-fixed rat brain. Briefly, a rat was euthanized prior to the experiment with pentobarbital, transcardially perfused with phosphate-buffered saline (PBS) followed by a 4% paraformaldehyde (PFA) solution. The brain was then extracted and stored in 4% PFA in a fridge at 4 degrees Celsius for 24 hours. The brain was washed with PBS for at least 48 hours before imaging to minimize relaxation-effects induced by the fixative (Birkl et al., 2016). The brain was subsequently placed in a plastic cylinder filled with Fluorinert (Sigma Aldrich, Lisbon, Portugal).

*MRI experiments*

Experiments were performed on a 9.4 T Bruker Biospec (Bruker, Karlsruhe, Germany) interfaced with an Avance IIIHD console and equipped with a single-channel volume coil. Remmi sequences ([Remmi](Remmi)) were used to acquire 3D gradient-recalled multi-echo images (MGE) and 3D dMRI images. For all acquisitions, repetition time was kept at 250 ms and the flip angle at 45 degrees. The Field-of-View (FOV) for these 3D acquisitions was 22.5×15.0 ×16.5 mm$^3$, matrix size 150×100×110 which resulted in an isotropic resolution of (150 μm)$^3$. For the MGE, the echo times were 4, 8.5,…, 26.50 ms, while dMRI was acquired at 20 ms. One average was acquired for the MGE and dMRI leading to an SNR around 40 in WM and 50 in GM for MGE, and an SNR around 4 in WM and 2 in GM for dMRI acquired with a b-value of 8 ms/μm$^2$ and along 75 directions. Diffusion times for the dMRI experiment was δ/Δ =7/9 ms. The sample was scanned at room temperature. Acquisition time was 45 minutes for MGE and 17 hours for dMRI. MGE was acquired at 5 different orientations described by yaw = 90, 45, 0, 0, 90 degrees and pitch = 0, 0, 0, 45, 45 degrees. FOV was only permuted when yaw or pitch was 90 degrees to keep the longest dimension parallel to the sagittal direction of the brain.

*Data processing*

MGE and dMRI processing was done as described in the manuscript. Images were further co-registered using an affine transformation to the brain positioned with yaw = 0 and pitch = 0, as this is the orientation where dMRI was acquired and the fODF was estimated. The rotation matrix from each co-registration was used to determine the direction of external field **B$_0$** to the brain. Susceptiblity fitting was done using Equations (8) and (9) corresponding to QSM and QSM+ (with mesoscopic frequency shift), respectively. We estimated the susceptibility for QSM and QSM+ for every direction using the LSMR algorithm. Tikhonov regularization was applied for each orientation and fitting algorithm based on L-curve optimization. Lastly, we estimated the susceptibility for QSM and QSM+ including all directions, corresponding to the COSMOS method.

# Results

*Multi orientation QSM (COSMOS)*

Figure S7 plots the residuals $\delta\overline{\Omega}_{MRI}$ from COSMOS and a voxel-by-voxel comparison of the values between QSM and QSM+. Figure S8 shows the susceptibility fits for QSM and QSM+. Here we find a lower residual for QSM+ compared to QSM, i.e., when we account for the mesoscopic frequency shift, and the variance $\sigma^2_{\hat{\mathbf{B}}}(\delta\overline{\Omega}_{MRI})$ seems visually less biased in the anterior commissure, where the largest (most anisotropic axons) mesoscopic frequency shift was found. This demonstrates the difference between QSM and QSM+ when the ill-posed dipole inversion is overdetermined and does not corrupt fitting performance.

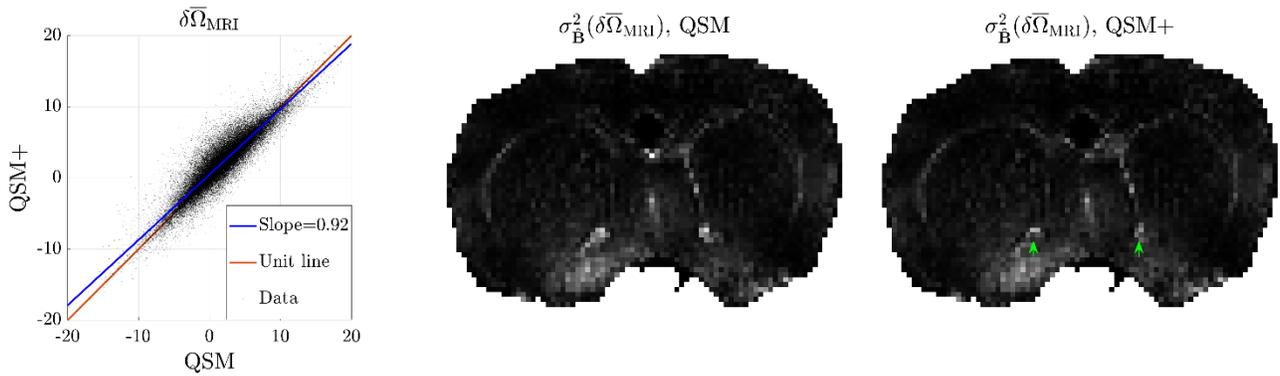

**Figure S7 – COSMOS Susceptibility fitting of rat brain at 150 µm isotropic resolution:** The plot to the left show voxel-by-voxel comparison of the residuals $\delta\overline{\Omega}_{\mathrm{MRI}}$ for fitting including all orientations. The red line corresponds to the unit line, while the blue shows a linear fit, with slope below 1, indicating lower residuals with QSM+. $\sigma^2_{\hat{\mathbf{B}}}(\delta\overline{\Omega}_{\mathrm{MRI}})$ shows the variance in the residuals for a coronal slice of the rat brain in the anterior part of the brain.

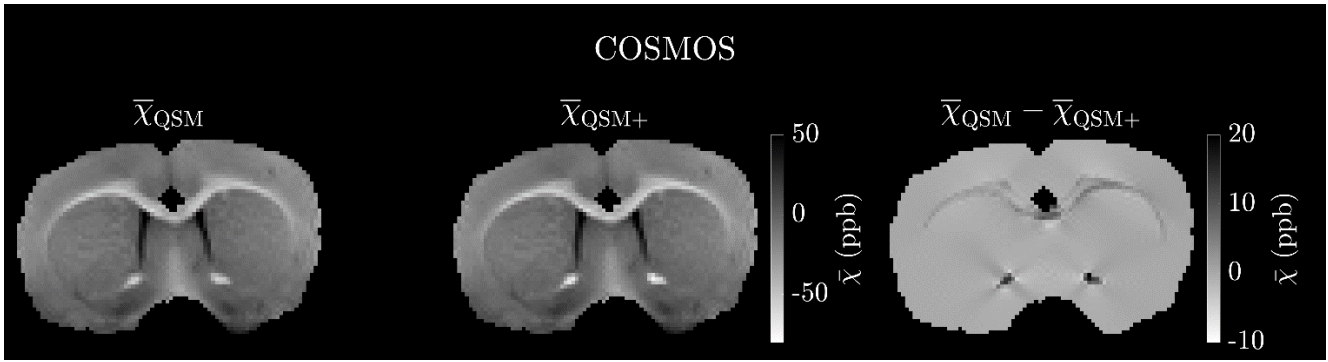

**Figure S8 – COSMOS Susceptibility maps of rat brain at 150 µm isotropic resolution:** coronal slices from the anterior part of the brain are shown. $\bar{\chi}_{\mathrm{QSM}}$ corresponds to zero mesoscopic contribution (conventional COSMOS), and $\bar{\chi}_{\mathrm{QSM+}}$ includes a non-zero mesoscopic contribution calculated using this method.

*Single orientation QSM*

Figure S9 shows the residual in Larmor frequency $\delta\overline{\Omega}_{\mathrm{MRI}}$ for each orientation, specified in the title. The figures are plotted voxel-by-voxel to compare QSM (x-axis) against QSM+ (y-axis). We find that the slope is slightly less than one, indicating a lower residual with QSM+. Figure S9 also illustrates the variance in $\sigma^2_{\hat{\mathbf{B}}}(\delta\overline{\Omega}_{\mathrm{MRI}})$ and $\sigma^2_{\hat{\mathbf{B}}}(\delta\overline{\chi})$. Only a slight improvement is found with QSM+. Nevertheless, the susceptibility found via QSM and QSM+ are different. Importantly, QSM+ for a single direction predicts a more negative magnetic susceptibility in WM, in agreement with the COSMOS QSM+ fit.

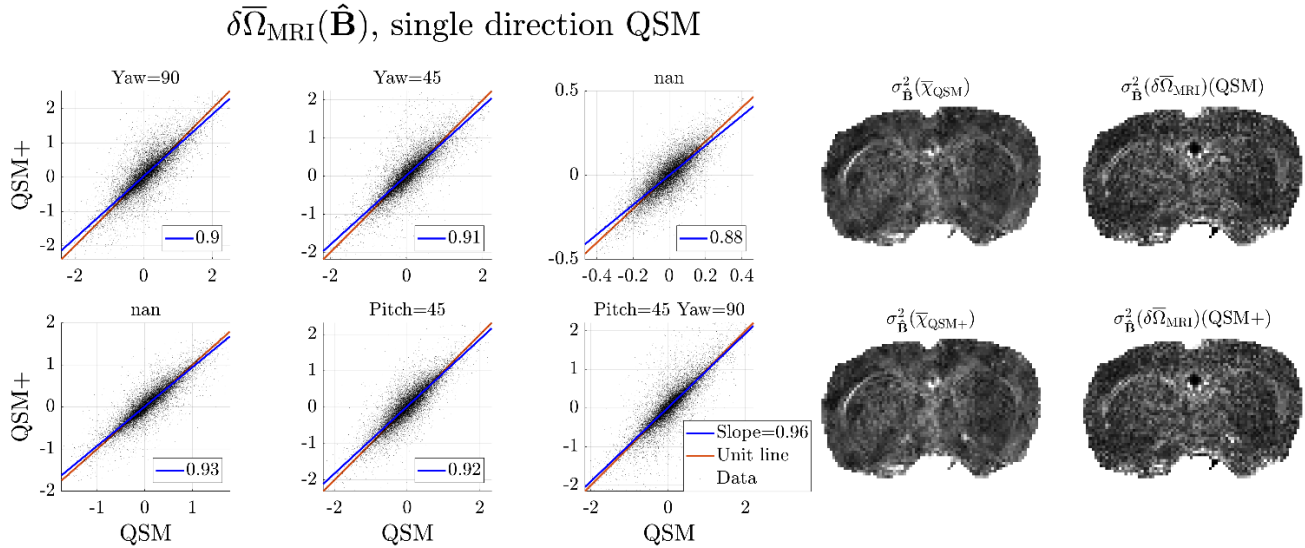

**Figure S9 - Susceptibility fitting of rat brain at 150 μm isotropic resolution at 5 different orientations:** The plots to the left show voxel-by-voxel comparison of the residuals $\delta\overline{\Omega}_{MRI}$ for each sample orientation labeled in the title. Nan corresponds to no rotation (two individual experiments are shown), and here the field is along the sagittal orientation of the brain. The red line corresponds to the unit line, while the blue shows a linear fit, with slope slightly below 1, indicating lower residuals with QSM+. $\sigma_{\hat{B}}^2(\delta\overline{\Omega}_{MRI})$ and $\sigma_{\hat{B}}^2(\delta\overline{\chi})$ show the variance in the residuals and susceptibility fits, respectively, for a coronal slice of the rat brain in the anterior part of the brain.